\documentclass[twocolumn,prb,showpacs,superscriptaddress,preprintnumbers,amsmath,amssymb]{revtex4}
\usepackage{graphicx}
\usepackage{mathptmx}
\usepackage{verbatim}
\usepackage{graphicx}
\usepackage{amsmath}
\usepackage{amssymb}
\usepackage{mathdots}
\usepackage{amsfonts}
\usepackage{mathrsfs}
\usepackage{amsmath}
\usepackage{color}
\usepackage{graphicx}
\usepackage{bm}    
\usepackage{amssymb}
\usepackage{xspace}
\usepackage{dcolumn}   

\begin{document}
\title{Origin of band gaps in graphene on hexagonal boron nitride}
\author{Jeil Jung}
\email{jeil.jung@gmail.com}
\affiliation{Graphene Research Centre and Department of Physics, National University of Singapore, 2 Science Drive 3, 117551, Singapore}
\author{Ashley M. DaSilva}
\affiliation{The University of Texas at Austin, Austin, Texas 78712, USA }
\author{Allan H. MacDonald} 
\affiliation{The University of Texas at Austin, Austin, Texas 78712, USA }
\author{Shaffique Adam}
\email{shaffique.adam@yale-nus.edu.sg}
\affiliation{Graphene Research Centre and Department of Physics, National University of Singapore, 2 Science Drive 3, 117551, Singapore}
\affiliation{Yale-NUS College, 6 College Avenue East, 138614, Singapore}



\begin{abstract}
Recent progress in preparing well controlled 2D van der Waals heterojunctions 
has opened up a new frontier in materials physics.  In this paper we address the intriguing 
energy gaps that are sometimes observed when a graphene sheet is placed on 
a hexagonal boron nitride substrate, demonstrating that they are produced by an interesting 
interplay between structural and electronic properties, including electronic many-body exchange interactions.  
Our theory is able to explain the observed gap behavior by accounting first for the structural relaxation of graphene's carbon atoms 
when placed on a boron nitride substrate and then for the influence of the substrate on low-energy 
$\pi$-electrons located at relaxed carbon atom sites.     
The methods we employ can be applied to many other van der Waals heterojunctions.
\end{abstract}

\maketitle

\section{Introduction}
\begin{figure}
\begin{center}
\includegraphics[width=8.4cm,angle=0]{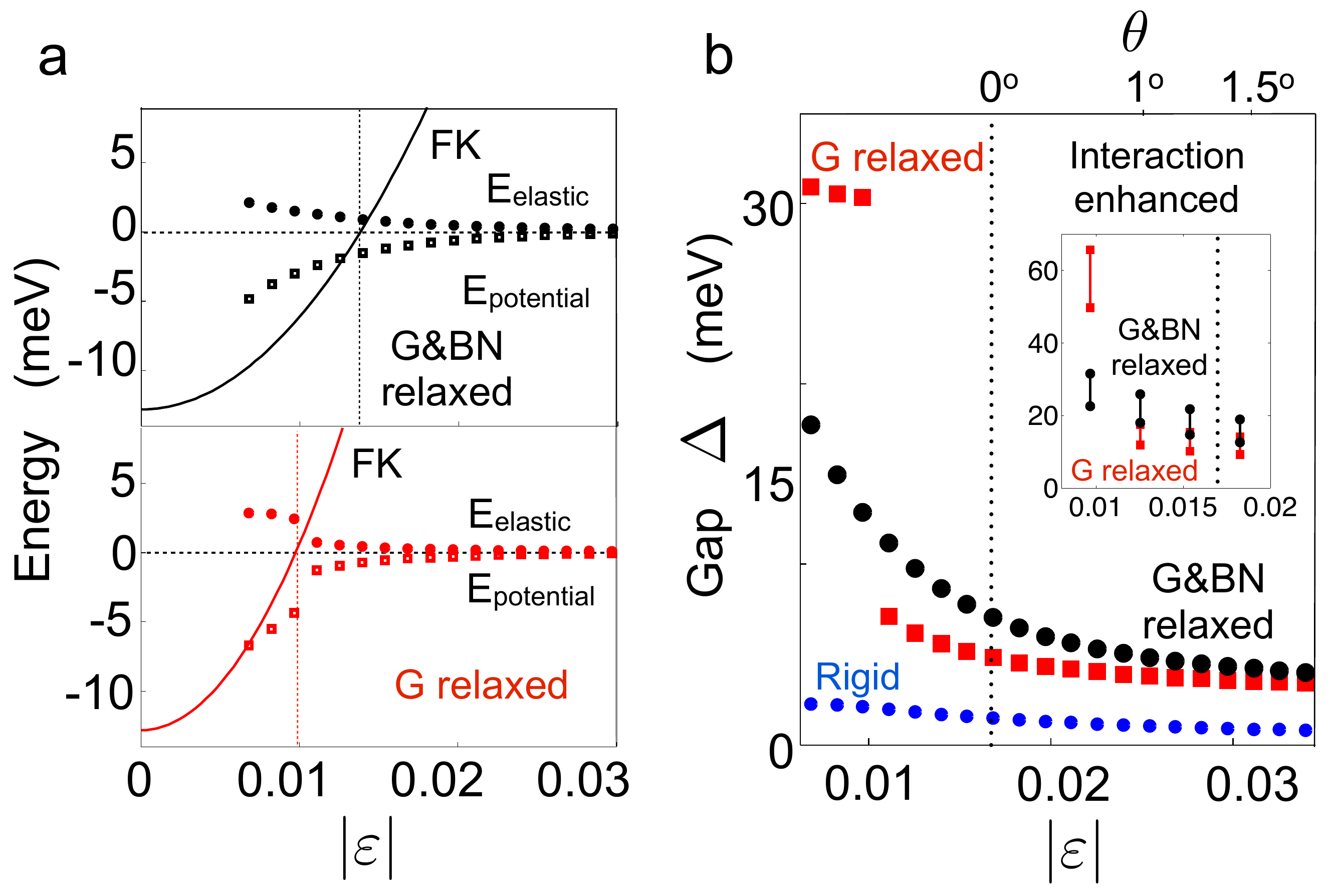}
\end{center}
\caption{{\bf Relaxation strains and band gaps of graphene on BN.}
{\bf a.} Relaxation strain elastic and potential energies for orientation aligned graphene on BN as a function of $\epsilon$ the 
relative lattice constant difference.  The black lines illustrates the case in which only carbon atom positions are allowed to 
relax (black) whereas the red curve is for the case in which both G and BN layer atoms are allowed to relax.
The parabolic curve labelled FK (Frenkel-Kontorova) plots the energy difference between an 
undistorted graphene sheet and one with a lattice that has expanded to be commensurate with that of the substrate
that is discussed in the text.  $E_{\rm elastic}$ and $E_{\rm potential}$ are respectively the elastic energy cost and the potential
energy gained by straining  both graphene and BN (black) layers, and the graphene (red) only while keeping the moir\'e 
lattice constant fixed.  
We use $\epsilon = -0.017$ for graphene on BN in the absence of graphene lattice expansion. 
{\bf b.} Energy gaps including strain effects {\it vs.} $\epsilon$ when graphene and BN layers are allowed to relax (black)
and when only graphene atoms are allowed to relax (red), when the layers are held rigid at 3.4~$\AA$ separation (blue).
and when electron-electron interactions are also included (inset).
The interaction enhanced gaps are bracketed by Hartree-Fock calculations that use 
dielectric constants of 2.5 and 4 to account for screening effects. 
The $\theta$ label indicates the one-to-one relation with $l_M$ 
when we fix $\left| \varepsilon \right| = 0.017$ and provides an approximate representation
of the twist angle dependence.
}
\label{figure1}
\end{figure}

Recent progress in preparing vertical heterojunctions of graphene (G)  and hexagonal boron nitride (BN) using 
either transfer~\cite{dean_seminal} or growth techniques~\cite{growthgbn} has opened a new frontier for exploring both fundamental 
physics~\cite{jarillo,kimhofstadter,geimhofstadter} and new device geometries~\cite{geimvertical}. 
Experiments have made it clear that graphene on BN is very flat and that its low-energy electronic states are often very weakly perturbed by the substrate~\cite{dean_seminal}.  However when the honeycomb lattices of graphene and BN are close to orientational alignment, the electronic coupling strengthens and is readily observed~\cite{leroy}.   The source of this variability in behavior is clearly related to variability in structure.  For example, although {\it ab initio} theory \cite{kelly1,moirebandtheory} predicts substantial gaps $\sim50$ meV when the two honeycomb lattices are identical, any incommensurability due to misorientation or lattice constant mismatch drastically reduces electronic coupling giving vanishingly small gaps~\cite{ortix}.  In this article we show that the large gaps observed~\cite{jarillo} 
at the Fermi level of neutral graphene sheets that 
are nearly rotationally aligned with a BN substrate are not due solely in terms of 
the relative orientation-dependent moir{\'e} pattern, but require in addition
both orientation-dependent structural relaxation of the carbon atoms, as suggested by recent experiments~\cite{geimgap},
and non-local many-body exchange interactions between electrons.
Our theory involves two elements: i) structural relaxation 
due to interactions between G and the BN substrate and ii) an effective Hamiltonian for graphene's $\pi$-electrons 
which includes a substrate interaction term that is dependent on the local coordination between
graphene and BN honeycombs.  
Our main results are summarized in Fig.~\ref{figure1}
where we show that atomic relaxation leads to substantially enhanced gap.
The band gap for rotationally aligned layers is only $\sim 1$ meV when the honeycomb lattices are held rigid,
but increases to $\sim$7 meV when relaxation is allowed.  
These gaps are further enhanced to $\sim20$ meV, in reasonable agreement with experiment,
when we also account for electron-electron interactions.   
Moreover, unlike other proposed mechanisms for band gaps in graphene \cite{mindthegap}, 
ours does not degrade the mobility of graphene.

\section{ Moir\'e patterns and strains} 
The $\pi$-electron Hamiltonian of G/BN can be expressed as the sum of the 
continuum model Dirac Hamiltonian of an isolated flat graphene sheet, in which the honeycomb sublattice
degree-of-freedom appears as a pseudo spin, 
and a correction from
the interaction with the BN substrate~\cite{leroy,moirebandtheory,ortix,falko1}.  We employ an approach in which the correction is given by a sub-lattice dependent but spatially local operator $H_{M}(\vec{d})$ derived from 
{\it ab initio} theory~\cite{moirebandtheory} that depends on the local alignment between G and BN honeycomb lattices $\vec{d}$.
This pseudo-spin dependent operator that gives rise to the moir\'e superlattice Hamiltonian
is accurately parameterized in Ref.~[\onlinecite{moirebandtheory}].
(An alternate parameterization which allows spatial variation in the interlayer separation is discussed in the 
appendix.)

When both G and BN form rigid honeycomb lattices
\begin{equation} 
\vec{d}(\vec{r}) \to \vec{d}_{0}(\vec{r}) \equiv  \epsilon \vec{r} + \theta \hat{z} \times \vec{r},
\end{equation} 
where $\epsilon$ is the difference between their lattice constants, $\theta$ is the difference in their orientations,
and $\hat{z}$ is the direction normal to the G sheet.  
The two layers establish a moir\'e pattern in which equivalent alignments 
repeat periodically on a length scale that, when $\epsilon$ and $\theta$ are small,
is long compared to the honeycomb lattice constant. 
(The moir{\'e} lattice vectors $\vec{L}_M$ solve $\vec{d}(\vec{r}+ \vec{L}_M) = \vec{d}(\vec{r}) +\vec{L}$ 
where $\vec{L}$ is a honeycomb lattice vector.)  
Since $H_{M}(\vec{d})=H_{M}(\vec{d}+\vec{L})$, the substrate interaction Hamiltonian has the 
periodicity of the moir\'e pattern.    

When the honeycomb lattices of the G and BN layers are allowed to relax, $\vec{d}(\vec{r})$ is no longer 
a simple linear function of position.  We write   
\begin{eqnarray}
\label{dvector}
\vec{d}(\vec{r}) = \vec{d}_0(\vec{r}) + \vec{u}(\vec{r}) + (h_0+h(\vec{r}) ) \, \hat{z}
\label{dofr}
\end{eqnarray}
where $h_0$ is the mean separation between G and BN planes and 
the in-plane and vertical strains, $\vec{u}(\vec{r})$ and $h(\vec{r})$, also have the moir\'e pattern periodicity.  If G/BN systems achieved
thermal equilibrium $\epsilon$, $\theta$, $\vec{u}(\vec{r})$ and $h(\vec{r})$ would be determined by 
minimizing free energy with respect to the positions of atoms in the G layer and in the BN 
layers close to the surface of the substrate.  Evidently this is not the case since 
the observed value of $\theta$ varies in an irreproducible fashion.  In the following we take the view that because the 
thermodynamic bias favoring a particular value of $\theta$ is weak, its observed value is 
fixed by transfer kinetics.  Similarly the value of $\epsilon$, which can be adjusted only by atomic 
rearrangements on long length scales, is also likely determined by kinetics and not by 
equilibrium considerations.  On the other hand, given values for $\epsilon$ and $\theta$ 
minimizing energy with respect to local strains $\vec{u}(\vec{r})$ and $h(\vec{r})$ require only local atomic arrangements.
We therefore view $\epsilon$ and $\theta$ as experimentally measurable system parameters.
In practice $\epsilon$ is close to the undisturbed relative lattice constant difference whereas 
$\theta$ varies widely.  The ratio of the honeycomb lattice constant to the moir\'e pattern lattice constant $l_M$ 
is $a/l_M =  (\varepsilon^2 + \theta^2)^{1/2}$.  
For given values of $\theta$ and $\epsilon$, $H_{M}(\vec{d}(\vec{r}))$ is dependent on
strains because of their contribution to Eq.~\ref{dofr}.  The strains must therefore be calculated
first in order to fix the $\theta,\epsilon$-dependent $\pi$-band Hamiltonian of G/BN.  
As a side remark, we note that $\vec{d}(\vec{r}) = (a / \l_M) \vec{r}$ is a convenient approximation
for the coordination vector that can account for the twist angle dependence through the magnitude of $l_M$
but ignores the variations in the shape of the moire pattern.

\begin{figure}
\begin{center}
\includegraphics[width=8.2cm,angle=0]{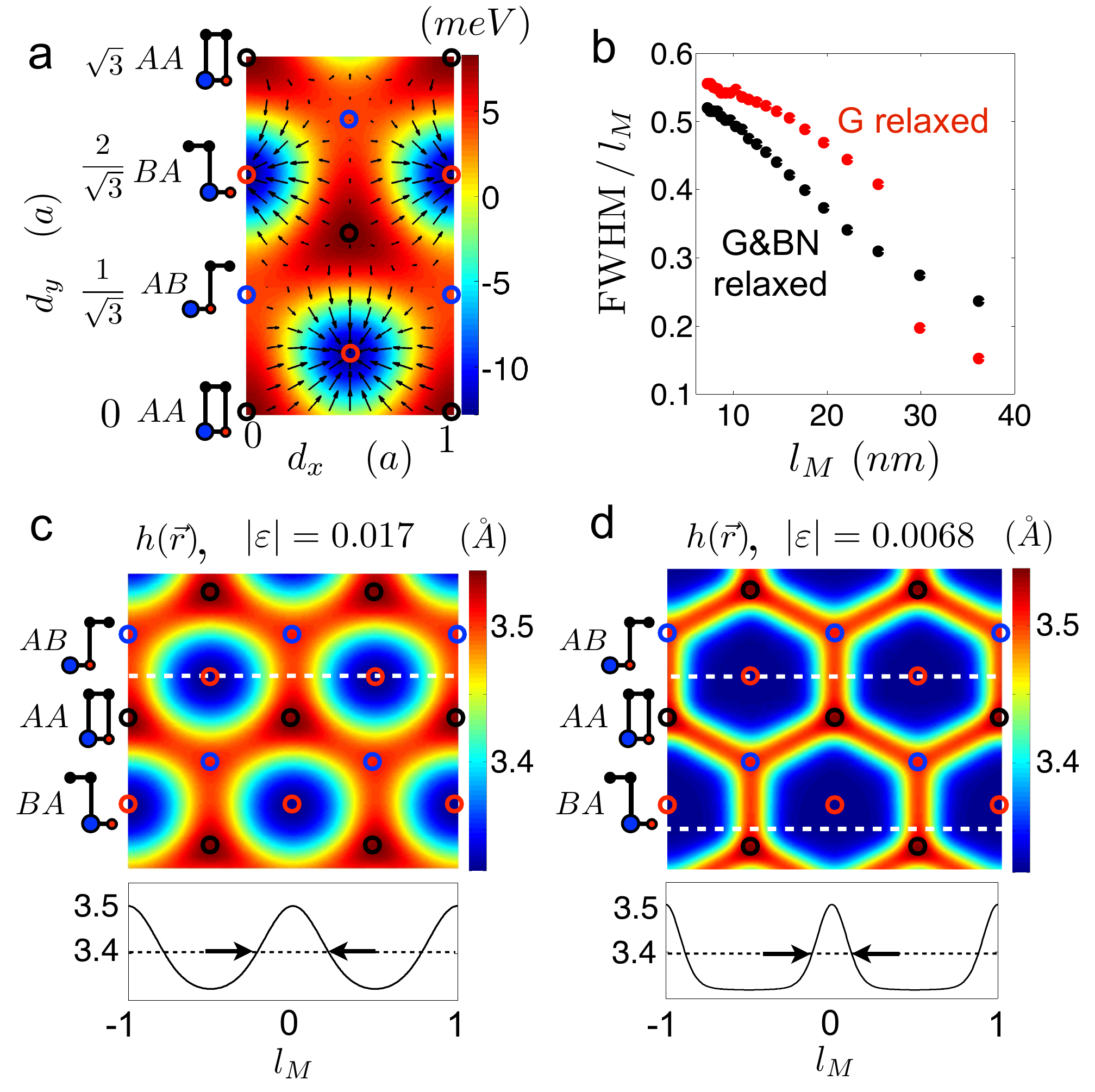}
\end{center}
\caption{
{\bf Relaxation strain and degree of commensuration as a function of moire pattern lattice constant.}
{\bf a}. Substrate interaction energy $U(\vec{d})$ per unit cell area as a function of stacking coordination $\vec{d}$. 
The arrows indicate the magnitudes and directions of substrate interaction forces 
$\vec{F} = -\vec{\nabla}_{\vec{d}} U$ which drive atoms toward local BA coordination.  
The  stacking arrangement cartoons use blue for boron, red for nitrogen, and black for carbon.
{\bf b}. Width of the distribution of carbon atom displacements (FWHM) as a function of moir\'e pattern
lattice constant at $\theta=0$.  
The typical displacement varies from $\sim 5$ to $\sim 8$~nm 
when the moir\'e pattern lattice constant varies by a factor of four.
{\bf c}. Vertical strains for $ \left| \varepsilon \right| = 0.017$ and {\bf d}. for $\left| \varepsilon \right| = 0.0068$.
Note that the vertical strain and the substrate interaction have similar spatial maps.  
The lower panels plot the height variation along the dashed lines of the upper panels for G\&BN relaxed geometries.  
The maps for the elastic and substrate interaction energies are discussed in the appendix. 
}
\label{figure3}
\end{figure}

Like the $\pi$-electron Hamiltonian, the graphene sheet energy can be written as the sum of 
an isolated G layer contribution and a substrate interaction contribution that depends on the local band 
alignment $\vec{d}(\vec{r})$.  The substrate interaction $U(\vec{d})$ is most attractive 
when half the carbon atoms are directly above boron atoms, and the centers of 
graphene's hexagonal plaquettes are directly above the nitrogen atoms (BA alignment).  
This alignment is energetically more stable than
one in which half the carbon atoms sit on top of nitrogen (AB alignment), or one in which 
all carbon atoms sit on top of either boron and nitrogen atoms (AA stacking).
By performing {\it ab initio} calculations for commensurate lattices 
we find that $U_{ BA} < U_{ AB} < U_{ AA}$.
The full dependence of $U$ on $\vec{d}$ is plotted in Fig. \ref{figure3}.

When $\epsilon$ or $\theta$ are non-zero, the substrate interaction forces 
plotted in Fig.~\ref{figure3} drive strains which attempt to match G and BN 
lattice constants locally and increase the sample area that is close to local BA coordination.
For a given value of $\epsilon$, G sheet lattice constant expansion near BA points must 
be compensated by lattice compression elsewhere.  
This kind of local expansion and compression of the graphene lattice within the moir{\' e} unit cell was
recently identified experimentally \cite{geimgap,alignment}. 

We determine the strains by minimizing the sum of the isolated graphene and substrate 
interaction energies.  For the long-period moir\'e lattices the graphene sheet energy is 
accurately parameterized in terms of its elastic constants.  The competition between isolated graphene and substrate
interaction energies can then be understood by comparing the energies of the configurations in which the two terms are 
minimized separately.  The substrate interaction energy is minimized by maintaining perfect BA alignment everywhere 
and therefore establishing commensurability between the BN and G lattices.  Because the lattice 
constants of BN and G differ, this arrangement has an elastic energy cost in the graphene sheet.  
After an elementary calculation we find that the total energy per area is  
\begin{equation} 
e_{BA} = U_{BA}/A_0 + 2 (\lambda + \mu) \epsilon_{0}^2 
\label{latticematch}
\end{equation} 
where $\lambda$ and $\mu$ are elastic constants, $\epsilon_{0}$ is the relative difference between BN and G lattice constants,
and $A_0$ is the unit cell area of graphene.
The elastic energy, on the other hand, is minimized by keeping the graphene sheet lattice constant at its isolated value.   In this 
configuration, because of the linear relationship between $\vec{d}$ and $\vec{r}$
the substrate interaction energy per unit area is equal to 
the average of $U(\vec{d})$ over $\vec{d}$:
\begin{equation}
e_{iso} = \overline{U}/A_0 >  U_{ BA}/A_0. 
\end{equation}  
As indicated in Fig.~\ref{figure1}, when our theoretical values for $U$ are 
combined with the elastic constants of a graphene sheet, the energy of the commensurate state 
is substantially lower.  However, Eq.~(\ref{latticematch}) overestimates the elastic energy cost of 
lattice matching between BN and G.  For example in the extreme case of a single BN layer, 
lattice matching can be achieved by adjusting the lattice constants of each layer toward their mean value, 
approximately reducing the 
required strains by a factor of 2.  
In this case, the incommensurate structure 
still has lower energy, but the difference is smaller.  We conclude that when they are orientationally aligned,
the interaction between a G sheet and a BN sheet is nearly strong enough to favor lattice matching. 

G/BN is close enough to an incommensurate to commensurate transition that 
substantial strains can be driven by substrate interactions.  
Indeed we find by explicit energy minimization that both vertical and horizontal strains can assume 
values large enough to introduce changes in the electronic structure. 
We determined these strains numerically for the case of a single-layer BN substrate
subject to a fixed periodic potential created by the layers underneath
using methods explained in the appendix.  
We find that strains in the graphene sheet are comparable as those in the BN layer.
Note that the atomic structure, and 
hence the $\pi$-band Hamiltonian, might therefore depend on the thickness of the BN and 
on other features that vary from one experimental study to another.
Similarly the addition of encapsulating layers can lead to reductions
in strains and hence gaps, as recently reported in Ref. \cite{geimgap}, 
although the gap can in principle persist.

\section{Strained moir\'e band Hamiltonian} 

Given $H_{M}(\vec{d})$ and $d(\vec{r})$, we obtain a sublattice-pseudospin dependent 
continuum Hamiltonian with the periodicity of the moir\'e pattern which is conveniently analyzed 
using a plane-wave expansion approach.  We write the full Hamiltonian in the form 
\begin{eqnarray} 
\label{hamiltonian}
\langle \vec{k}',s'| H | \vec{k},s \rangle &=& \delta_{\vec{k},\vec{k'}} \langle s' |H_{D}(\vec{k}) |s\rangle  +   \nonumber \\
&+&  \sum_{\vec{G}}  \langle s' |H_{M, \, \vec{G}}|s\rangle   \; \Delta\left(\vec{k'}-\vec{k}  -\vec{G} \right)  
\label{effham}
\end{eqnarray} 
where $H_{D}$ is the Dirac Hamiltonian and $H_{M, \,\vec{G}}$ is the Fourier transform over one period 
of the moir\'e pattern of $H_{M}(\vec{d}(\vec{r}))$, and $\vec{G}$ is a moir{\' e} pattern reciprocal lattice vector.  
In Eq.~(\ref{effham}), $\Delta(\vec{k})=1$ when $\vec{k}$ is a moir\'e pattern reciprocal lattice vector and 
zero otherwise.  

\begin{figure*}
\begin{center}
\includegraphics[width=17cm,angle=0]{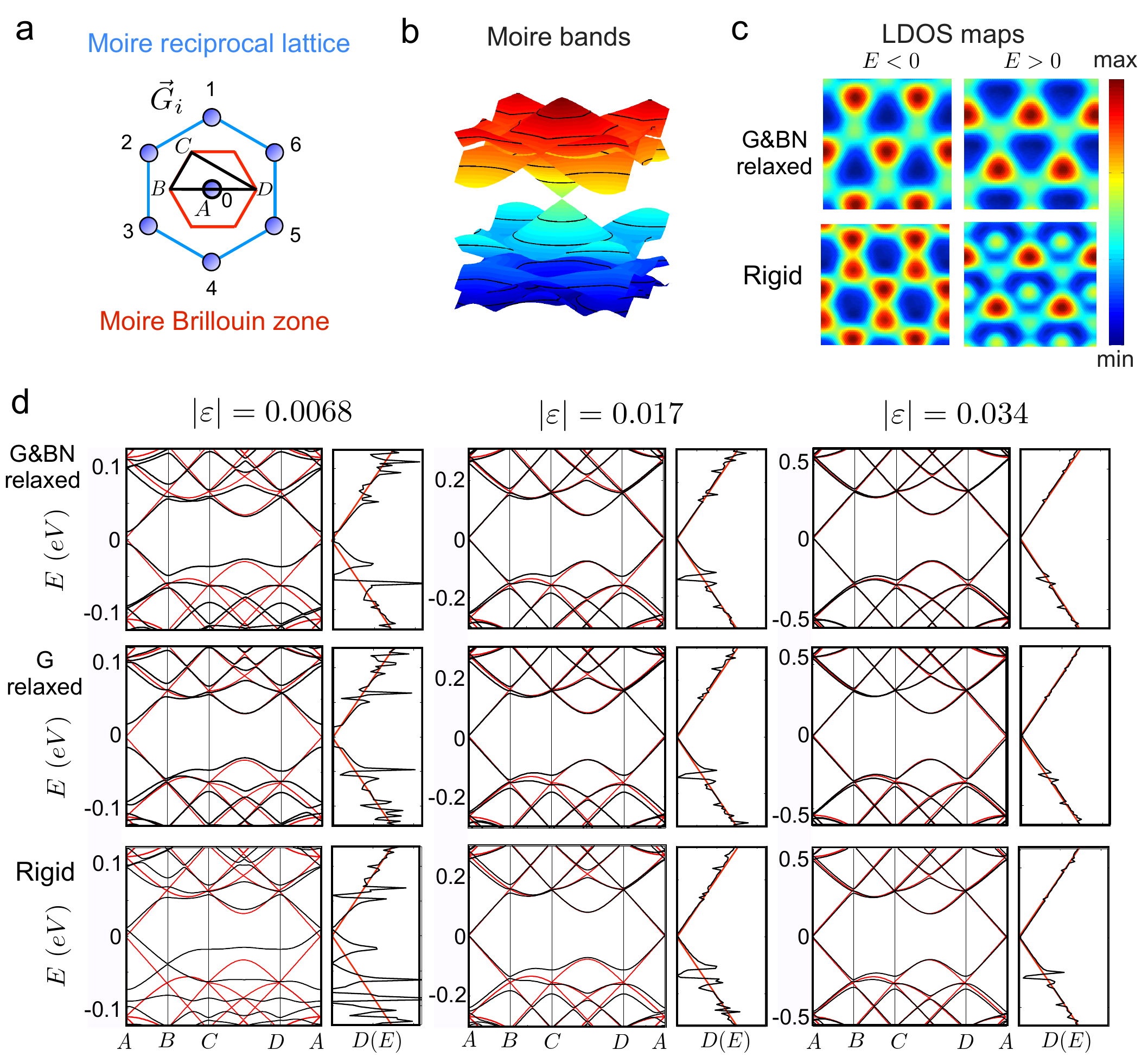}
\end{center}
\caption{{\bf Electronic structure of G/BN heterojunctions.}
{\bf a}. Schematic representation of the moir\'e Brillouin zone and the moir\'e reciprocal lattice vectors. 
{\bf b}. Three dimensional representation of the band structure in the moir\'e Brillouin zone showing superlattice Dirac point features.
{\bf c}. Local density of states (LDOS) maps near the charge neutrality Fermi energy for G\&BN relaxed and rigid lattice
structures at $\theta=0$ that show contrasts for electrons and hole carrier doping.  Lattice relaxation affects the LDOS maps. 
{\bf d}. Band structure and density of states for three different values of $\epsilon$ 
at $\theta=0$ allowing G\&BN relaxation, G-relaxation only, and with no relaxation.
In-plane lattice relaxation leads to sizeable band gaps in the limit of long moir\'e periods.
}
\label{figure2}
\end{figure*}

The electronic structures implied by the Hamiltonian in Eq.~(\ref{hamiltonian})
for rigid lattices, for graphene relaxation only, and for mutual G and BN relaxation are 
compared in Fig.~(\ref{figure2}).  
These results demonstrate that the electronic structure, and the gap at neutrality in particular,
depend sensitively not only on $\theta$ and $\epsilon$ but also on the strains.  
Sizable band gaps appear at the neutral system Fermi level  
only when in-plane relaxation strains $\vec{u}(\vec{r})$ are allowed.

\section{Physics of the Gaps} 
Several potential mechanisms of gap formation in neutral graphene  
have been discussed in the literature including antidots~\cite{pedersen},
combinations of periodic scalar and vector fields~\cite{snyman,guinealow},
and zero-line localization~\cite{kindermann}.  Our approach allows for a 
simple classification based on the Fourier expansion of $H_{M}$.
We will discuss leading contributions to the gap at neutrality in terms of the expansion of each 
moir\'e pattern Fourier component of $H_{M}$ into four sublattice Pauli matrix components.
We start with the $\vec{G}_0=0$ Fourier component, {\it i.e} with the spatial average of 
$H_{M}$.  In the absence of relaxation, $H_{M, \vec{G}_0=0}=0$ 
because the average of $H_{M}(\vec{d})$ is zero~\cite{moirebandtheory} and $\vec{d}$ in this case is a 
linear function of $\vec{r}$.   (We neglect an irrelevant contribution proportional to 
the identity sublattice Pauli matrix $\tau^{0}$.)  When $\vec{d}$ is a non-linear function 
of $\vec{r}$, however, the spatial average of Hamiltonian contributions which are 
sinusoidal functions of $\vec{d}$ do not vanish.  Among these, the term 
proportional to $\tau^{0}$ is an irrelevant constant, and the terms proportional to 
$\tau^{x}$ and $\tau^{y}$,   
often interpreted in Dirac models as effective vector potentials,
simply shift band crossings away from zero momentum.
(In the continuum Dirac model of graphene, momentum is measured away from the Brillouin-zone corners.) 
However, the $\vec{G}=0$ term proportional to $\tau^{z}$ produces a gap 
\begin{eqnarray}
\Delta_0  = 2 H_{M,\vec{G}_0=0}^{z} = H_{M,\vec{G}_0=0}^{AA} - H_{M, \vec{G}_0=0}^{BB}. 
\label{zerothorder}
\end{eqnarray} 
Physically this gap appears simply because the average site energy is 
different on different honeycomb sublattices.

The leading contributions to the gap from $\vec{G} \ne 0$ terms in $H_{M}$ 
are more subtle and appear at second order in perturbation theory.  A perturbative treatment is in fact 
valid in practice because it turns out that $\hbar \upsilon |\vec{G}|$ is substantially larger than $H_{M, \vec{G}}$.
Applying degenerate state perturbation theory we obtain the following expression for 
their contribution to the effective $2 \times 2$ sublattice Hamiltonian at the Dirac point:
\begin{equation}\label{Heffective}
H_{\rm eff}=H_{M, \vec{G}_0=0} -\sum_{\vec{G}\ne 0} H_{M, \vec{G}} \, H_{\vec{G}}^{-1}\, H_{M, \vec{G}}^{\dagger},
\end{equation} 
where $H_{\vec{G}}=\hbar \upsilon \tilde{G} \cdot \vec{\tau}$ ($\upsilon$ is the Dirac velocity and $\vec{\tau}$ is the vector of Pauli matrices,) ignoring the $\vec{k}$ dependence close to the Dirac point which will be higher order. 
Note that $H_{M, \vec{G}}$ connects the $\vec{k}$ and $\vec{k} + \vec{G}$ 
blocks of the plane-wave expansion moir\'e band Hamiltonian.  
Because only the term proportional to $\tau^{z}$ can produce a gap at 2nd order,
it is instructive to decompose $H_{\rm eff}$ into Pauli matrix contributions.
\begin{equation}
H_{\rm eff}= \sum_{\alpha=0,x,y,z} H_{\rm eff}^{\alpha} \; \tau^{\alpha}.
\end{equation}
Note that higher order terms proportional to $\tau_{x}$ and $\tau_{y}$ may in principle contribute to the gap, but we find them to be negligible.
We have derived analytic expressions for $H_{\rm eff}^{\alpha}$ in terms of the
Pauli matrix decomposition of $H_{M,\vec{G}_i}$ which are discussed in detail 
in the appendix.  We find that although the $\vec{G}=0$ 
contribution to the gap is always larger, the $\vec{G} \ne 0$ 
contributions are not negligible.  Both the difference in the spatial average of 
sub band energies and the detailed form of the full substrate interaction 
Hamiltonian play a role in determining the size of the gap at neutrality,
and both are sensitive to the detailed structure of the lattice relaxation strains.

In graphene non-local exchange interactions are expected to enhance gaps~\cite{physicascripta,song,kellygap} 
at neutrality produced by sublattice-dependent potentials.  
We have performed plane-wave-expansion self-consistent Hartree-Fock calculations
in which Coulomb interactions are added to the moir\'e band Hamiltonian we have discussed.
The calculations were performed using effective dielectric constants
bracketing the expected values between $\varepsilon=2.5$ and $\varepsilon_r = 4$.
When all effects are included
we find band gaps $\sim20$~meV, as shown in the inset of Fig. \ref{figure1}. 
The values chosen for $\varepsilon$ partly account for dielectric screening by the 
substrate and partly accounts for dynamic screening effects in the same 
spirit as in the screened exchange functionals used in density-functional theory.  
We have previously used a similar dielectric constant of $\varepsilon_r = 4$ to successfully predict 
spontaneous band gaps $\sim 50$~meV in ABC trilayer graphene \cite{trilayergap}.
Further details of the Hartree-Fock theory in moire superlattice bands will be presented elsewhere \cite{moireHF}.

\section{Discussion}

We have derived a $\pi$-band continuum model Hamiltonian intended to describe states 
near the Fermi level of G/BN and used it to address the energy gaps often observed 
in neutral graphene when it is nearly aligned with a BN substrate.  In this theory the 
interaction of $\pi$-band electrons with the substrate is described by a local but 
sublattice dependent term $H_{M}$ that is dependent on the local relative displacement of the 
graphene sheet and substrate honeycomb lattices, $d(\vec{r})$.  When neither the G sheet's carbon atoms nor 
the boron and nitrogen atoms in the substrate are allowed to relax, $d(\vec{r})$ is a linear function of position 
because of the difference between the lattice constants $\epsilon$ and because of difference in
orientations specified by a relative angle $\theta$.  The gap produced by substrate interactions 
in the absence of relaxation reaches its maximum at $\theta=0$, 
but is never larger than a few meV and too small to explain experimental observations.  
Only by allowing the carbon and substrate atoms to relax we can explain the much
larger experimental gaps.  

The moir\'e pattern formed by graphene and a BN substrate is characterized in the 
first place by the lattice constant difference $\epsilon$ and by the relative orientation angle $\theta$. 
These two quantities can be changed only by collective motion of many atoms.
We take the view that because of large barriers and weak thermodynamic drivers
these two macroscopic variables are not in practice relaxed to equilibrium values.
We therefore view them as observables that characterize particular G/BN systems and calculate 
relaxation strains and $\pi$-band electronic structure as a function of $\epsilon$ and $\theta$, and hence as a 
function of moir\'e pattern period.  The explicit calculations reported on in this paper are for $\theta=0$,
the orientation which leads to large experimental gaps.  

To account for relaxation strains, we minimize the total energy with respect to carbon and 
substrate atom positions.  For this purpose we assume that the interaction energy $U$ between 
graphene and substrate is also a local function of $\vec{d}$ and obtain $U(\vec{d})$ 
from density functional calculations of commensurate structures.  
The strains minimize the total energy by increasing the number of carbon atoms that are on top of boron 
atoms and the number of hexagonal carbon atom plaquettes that are centered above nitrogen atoms.  
Our study emphasizes that atom relaxation in the BN sheets is as important for the 
electronic structure as atom relaxation in the graphene sheet.  Although only atom positions in the 
top BN sheet are important for electronic structure, these will be affected by interactions 
with atoms in remote layers.  We have performed calculations for two extreme cases, rigid 
BN atoms and a single-layer of BN in which atom positions relax to minimize total energy,
finding that relaxation increases the energy gap substantially. The physical origin of these gaps can be revealed by expanding the continuum model
$\pi$-band Hamiltonian in terms of Pauli-matrix pseudospin operators and in terms of 
moir\'e pattern reciprocal lattice vector components.  Because of the wide $\pi$-band width and 
the relatively short moir\'e periods, the contribution of each term in the Hamiltonian to the gap 
can be analyzed using leading order perturbation theory.  The $\vec{G} \ne 0$ terms which capture
detailed spatial patterns contribute at second order and are not negligible.  The largest contribution 
to the gap comes from the $\vec{G}=0$ term, which vanishes in the absence of 
lattice relaxation has a very simple interpretation.  Because of relaxation strains the average site 
energy in the carbon sheet is different for the two carbon atom sub lattices.  It is well known that 
this type of perturbation produces a gap at the Fermi level of a neutral graphene sheet. 
Surprisingly the gap is a substantial fraction of the gap of the same origin present in the 
commensurate BA aligned graphene on BN.  The gaps are therefore due to the contrast between the 
local classical physics of energy minimization with respect to atom position, and the wide $\pi$-bands 
and non-local quantum physics which forces the quantum wave functions to be smooth and sensitive mainly
to spatial averages over the moir\'e period.  When many-body interaction effects~\cite{physicascripta} are accounted for,
these gaps are enhanced to values that are consistent with experiment.
The approach described in this paper can be applied to other van der Waals materials 
which can form heterojunctions in which different layers have slightly different lattice 
constants or differ in orientation - such as transition metal dichalcogenide stacks.

\section{Methods}
The elastic energy functional was modeled using the Born-von Karman plate theory \cite{elasticitybook}.
Neglecting the small bending rigidity of graphene $\kappa = 1.6$~eV \cite{bendingstiffness}
the elasticity theory depends on the two Lam\'e parameters whose 
estimates for graphene from empirical potentials gives $\lambda = 3.25$ eV/$\AA^2$ and $\mu = 9.57$ eV/$\AA$ \cite{katsnelson}
in the low temperature limit, and for a single BN sheet we have used
$\lambda \sim  3.5  \,\, {\rm eV} \, \AA^{-2}$ and $\mu \sim 7.8 \,\, {\rm eV} \, \AA^{-2}$  \cite{sachs}
obtained averaging the LDA and GGA values.
The potential energy has been parametrized from the stacking-dependent and separation dependent 
energy curves in Ref. [\onlinecite{sachs}] calculated at the EXX+RPA level.
The scalar functions used to obtain the moir\'e superlattice pattern for the height and the displacement vectors
from their gradients have used $\Phi$ written as a Fourier expansion in $\vec{G}$ vectors as
\begin{eqnarray}
\Phi ( \vec{d} ) &=& \sum_{\vec{G}} C_{\vec{G}} \exp \left( - i \vec{G} \cdot \vec{d} \right)   \label{scalarpotential}   \\
&\simeq&  C_0 + C'_1 g(\vec{d}) + f_1(\vec{d},C_1,\varphi_1) +  f_2(\vec{d},C_2,\varphi_2)    \nonumber
\end{eqnarray}
where $C_{\vec{G}}$ is in general a complex number and we retain up to three nearest $\vec{G}$ vectors for 
the scalar field that preserves the symmetry of triangular lattices.
The parameters $C_0$, $C'_1$, $C_1$, $\varphi_1$, $C_2$ and $\varphi_2$ are real valued constants 
and we defined auxiliary functions $f$ and $g$ 
in terms of the triangular lattice  structure factors similar as those used in a general 
tight-binding model of graphene \cite{graphenetb}.
The Fourier expansion coefficients within the first shell consisting of $C'_0$ and the first shell $ f_1(\vec{d},C_1,\varphi_1)$ 
are often good representation of the solutions.
Further details on the calculation method and results of the elastostatic problem can be found in the appendix.
The self-consistency Hartree-Fock calculations were calculated using an effective relative dielectric constant of $\varepsilon_r = 4$ 
using 217 $k$-points in the moir\'e  Brillouin zone.

\section{Acknowledgments}
The work in Singapore was supported by National Research Foundation of Singapore under its Fellowship program (NRF-NRFF2012-01).  
The work in Austin was supported by the Department of Energy, Office of Basic Energy Sciences under contract DE-FG02-ER45118, 
and by the Welch Foundation grant TBF1473.  We acknowledge use of computational resources from the Texas Advanced Computing Center
and the high performance computing center at the Graphene Research Centre at the National University of Singapore. 
We acknowledge helpful discussions with Andre Geim and Gareth Jones and the 
assistance from Miguel Dias Costa for troubleshooting the parallel implementation of the Hartree-Fock calculations.

\section{Author contributions}
JJ and AMD executed research. 
All authors contributed in conceiving, discussing and preparing the manuscript. 

\section{Conflicts in financial interest}
The authors declare no conflicting financial interests.


\bigskip
\bigskip

{\large \bf Appendix.}
\appendix

In the following we supplement the information in the main text introducing:
the parametrization of the Hamiltonian in real-space,
the explicit form of the scalar fields with the moir\'e superlattice symmetry,
the elastic energy functionals, the parametrization of the potential energy,
the formulation and solutions to the elastostatic problem for the relaxed ground states,
and the contributions of the different pseudospin terms in the Hamiltonian to the 
primary Dirac point band gap, both numerically 
and analytically through second-order perturbation theory.

\section{Parametrization of the Hamiltonian}
The diagonal and off-diagonal elements of the Hamiltonian 
for a fixed interlayer separation distance can be written in the sublattice
basis in a manner similar to the pseudospin representation 
used in Ref. [\onlinecite{supp_moirebandtheory}], 
\begin{widetext}
\begin{eqnarray}
H_{ii}(\vec{K} :\vec{d}) &=&  2 C_{ii} {\rm Re}[ f(\vec{d})  \exp[i \varphi_{0}]  ]    \label{hofd} \\
H_{AB}(\vec{K} :\vec{d}) &=& 2 C_{AB} \, \cos( \frac{\sqrt{3} }{2} G_1 d_x) 
\left(  
\cos \left( \frac{G_1 d_y}{2}  - \varphi_{AB}  \right) 
+\sin \left( \frac{G_1 d_y}{2}  - \varphi_{AB} - \frac{\pi}{6}  \right)
\right)  
+ 2 C_{AB} \, \sin \left( G_1 d_y + \varphi_{AB} - \frac{\pi}{6}  \right)   \nonumber \\
&+& i \,2 C_{AB} \, \sin( \frac{\sqrt{3}  }{2} G_1 d_x ) 
\left(  
\cos \left( \frac{G_1 d_y}{2}  - \varphi_{AB}  \right)   
- \sin \left( \frac{G_1 d_y}{2}  - \varphi_{AB} - \frac{\pi}{6}  \right)
\right).    
 \nonumber
\end{eqnarray}
\end{widetext}
The out-of-plane $z$-axis layer separation dependence can be incorporated into the 
three main coefficients $C_{ii}(z)$ with an exponentially decaying behavior in the form
\begin{eqnarray} 
C(z) = C({z_0}) \exp(-B\cdot (z - z_0))
\end{eqnarray}
where $z_0 = 3.35 \, \AA$, and the three decay coefficients $B = 3.0, \, 3.2, \, 3.3 \,\, \AA^{-1}$
for each one of the terms of the Hamiltonian in the sublattice basis
were found fitting the $z$-dependence between 2.8 $\AA$ to 5 $\AA$,
where we use the parameters obtained from {\em ab initio} calculations
\begin{eqnarray}
C_{AA}(z_0) &=& -14.88 \,\, {\rm meV},   \quad   \varphi_{AA} = 50.19^{\circ}   \\
C_{BB}(z_0) &=& 12.09 \,\, {\rm meV},  \quad \varphi_{BB} =   - 46.64^{\circ} \\ 
C_{AB}(z_0) &=& 11.34 \,\, {\rm meV}, \quad \varphi_{AB} =  19.60^{\circ}
\end{eqnarray}
whose equivalent values in the pseudospin basis had been calculated previously \cite{supp_moirebandtheory}. 
The variation of the phase with $z$ shows a weak linear dependence and we can
approximate it as a constant value. 
The effects due to lattice relaxation can be conveniently incorporated when calculating the 
Fourier expansion of the above Hamiltonian 
by accounting for the in-plane displacement $\vec{u}(\vec{r}) = (u_x(\vec{r}),u_y(\vec{r}))$ 
in the stacking coordination vector $\vec{d}(\vec{r}) = \vec{d}_0(\vec{r}) + \vec{u}(\vec{r})$,
as explained in the main text, and the height $z = h(\vec{r})$ that represents the local distance of graphene to BN,
where $\vec{r}=(x,y)$ is a two-dimensional vector. 
Both the displacement vectors $\vec{u}(\vec{r})$ and the height maps $h(\vec{r})$ are 
assumed to respect the moir\'e periodicity and are therefore modeled from the
scalar fields that we introduce in the following section.

\section{Scalar fields for describing the moir\'e patterns}
\label{scalarpot}
\begin{figure}[htbp]
\begin{center}
\includegraphics[width=7cm,angle=0]{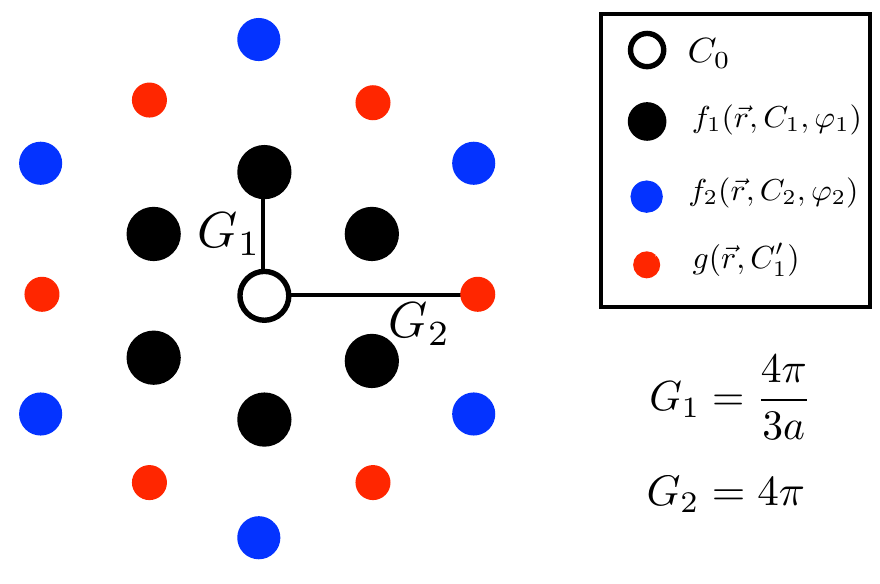}
\end{center}
\caption{
Representation of the different shells corresponding to the different structure factors 
$f_1$, $f_2$ and $g$ and the parameters that define the specific form of the scalar functions.
}
\label{landscape}
\end{figure}
In the main text we presented an approximation for a scalar field that varies smoothly in real space
that respects the symmetry of the triangular superlattice to use in the variational trial functions.
For brevity in notation here we use $(x,y)$ to indicate the $(d_x, d_y)$.
The specific form of the trial functions we use are given by
\begin{eqnarray}
\Phi\left( \vec{r} \right) &=& \sum_{\vec{G}} C_{\vec{G}} \exp \left( - i \vec{G} \cdot \vec{r} \right)  \label{scalarfield} \\
&\simeq&  C_0 + C'_1 g(\vec{r}) + f_1(\vec{r},C_1,\varphi_1) +  f_2(\vec{r},C_2,\varphi_2)     \nonumber
\end{eqnarray}
where the constants $C_{\vec{G}}$ are complex numbers.
The $f$ function 
\begin{eqnarray}
f_j (\vec{r},C_j,\varphi_j) &=& C_j \exp(i \varphi_j) \widetilde{f}_j (\vec{r})  + {\rm c.c.}
\end{eqnarray}
is defined in terms of the structure factors 
\begin{eqnarray}
\widetilde{f}_j (\vec{r}) &=& \exp (- i j G_1  y  )     \nonumber \\
 &+&  2 \exp ( i  j G_1  y / 2 ) \cos( j \sqrt{3} G_1  x / 2 )
\end{eqnarray}
where $G_1 = 4\pi/3a$, where $a$ is the real-space periodicity of the moir\'e superlattice and $j = 1, 2$.
These are momentum space analogues of the real space inter-sublattice hopping structure factors
in a honeycomb lattice \cite{supp_graphenetb}.
The explicit form of the functions defined along the symmetry lines $x = 0$ or  $y = 0$ can be obtained
from sums of 
\begin{eqnarray}
f_j(\vec{r}, y=0, C, \varphi) &=& 2 C \cos \varphi \left( 1 + 2 \cos( j \sqrt{3} G_1 x / 2 )\right)  \nonumber \\
f_j(\vec{r}, x=0, C, \varphi) &=& 2 C \cos \left( \varphi - jG_1 y  \right)    \nonumber \\
&+& 4 C \cos \left(  \varphi + jG_1 y / 2  \right).  
\end{eqnarray}
The analytical expression for the $g$ function shell contribution reduces to a simpler form
\begin{eqnarray}
g (\vec{r}) = 2 \cos \left( G_2 x \right) + 4 \cos \left( \sqrt{3} G_2 y / 2 \right) \cos \left( G_2 x / 2 \right),
\end{eqnarray}
where $G_2 = 4 \pi$,
that for the symmetry lines reduce to 
\begin{eqnarray}
g(\vec{r}, y = 0 ) &=& 2 \cos \left( G_2 x \right) + 4 \cos \left( G_2 x / 2  \right)  \nonumber \\
g(\vec{r}, x = 0 ) &=& 2  + 4 \cos \left( \sqrt{3} G_2 y / 2   \right). 
\end{eqnarray}
The vector fields such as in-plane forces, displacement vectors, and stresses 
can be obtained as gradients of the scalar potentials given by the above forms
that can preserve the symmetry of the triangular moir\'e superlattice.
The vector field that can be obtained from the gradient of the scalar field is
\begin{eqnarray}
\vec{\nabla} \Phi =  \vec{\nabla} f_1 + \vec{\nabla} f_2 + \vec{\nabla} g  
\label{vectorfield}
\end{eqnarray}
and can be obtained taking the respective partial derivatives.
Thus we have
\begin{eqnarray}
\vec{\nabla} f_j = C_j  \exp(i \varphi_j) \vec{\nabla} \widetilde{f}_j  + c.c.
\end{eqnarray}
where the partial derivatives of the constituent functions are given by 
\begin{eqnarray}
\partial_x \widetilde{f}_j(\vec{r})  &=& - j \sqrt{3}G_1 \exp(ij G_1 y / 2 ) \sin(j \sqrt{3}G_1 x / 2)    \nonumber  \\
\partial_y \widetilde{f}_j(\vec{r})  &=&  ijG_1 \left( - \exp(-ij G_1 y  )  \right.  \nonumber \\ 
&+& \left. \exp(ij G_1 y / 2 ) \cos(j \sqrt{3} G_1 x/2) \right).    
\end{eqnarray}
For the $g$ terms we have
\begin{eqnarray}
\partial_x g(\vec{r})  &=& - 2 G_2  \left( \sin (G_2 x)  + \cos(\sqrt{3}G_2 y/2) \sin(G_2 x/2) \right)     \nonumber   \\
\partial_y g(\vec{r})  &=&  - 2 \sqrt{3} G_2 \sin(\sqrt{3} G_2 y/2) \cos( G_2 x/2).   
\end{eqnarray}
Likewise higher order derivatives used in the stress tensors or the gauge fields can be evaluated analytically.
The pair of parameters $C$ and $\varphi$ for each $f_j$ function and 
the single parameter accompanying the $g$ function specify the variational space we used to minimize the energy functionals. 
Because the $f_1$ term captures the first harmonic contribution, the different variables such as $\vec{u}$, $\vec{h}$, $E_{pot}$ can 
be characterized in terms of just two parameters $C_1$ and $\varphi_1$,
or up to three when the average value of the origin $C_0$ is required.

\section{Elastic and potential energy functionals}

\subsection{Elastic energy functional}

The elastic energy for the Born-von Karman plate theory
 can be obtained in terms of the Lam\'e parameters since the small
 bending stiffness $\kappa$ for graphene or BN plays a negligible role \cite{supp_bendingstiffness}.
For graphene we use
$\lambda = 3.25  \,\, {\rm eV} \, \AA^{-2}$, $\mu = 9.57 \,\, {\rm eV} \, \AA^{-2}$ valid close to zero 0 K
have been estimated from empirical potentials \cite{supp_katsnelson}
and the for a single sheet of BN use the DFT estimates
$\lambda = 3.5  \,\, {\rm eV} \, \AA^{-2}$, $\mu = 7.8 \,\, {\rm eV} \, \AA^{-2}$
obtained averaging the LDA and GGA values.
The total elastic energy per superlattice area is given by  \cite{supp_elasticitybook,supp_guinearipples} 
\begin{eqnarray}
\label{elasticityeq}
E_{\rm elastic} &= &   \frac{\kappa}{2 A_M} \int_{A_M} d^2 \vec{r} \left[
    \nabla^2 h ( \vec{r} )
\right]^2  +  \\ &+ &
\frac{1}{2 A_M}\int_{A_M} d^2 \vec{r} \left\{   \lambda \left[
 u_{11} ( \vec{r} ) + u_{22} ( \vec{r} ) \right]^2   \right.  \nonumber \\
 &+&  \left.
2 \mu  \left[ u_{11}^2 ( \vec{r} ) + u_{22}^2 ( \vec{r} )  +  u_{12}^2 ( \vec{r} ) \right] \right\} \nonumber 
\end{eqnarray}

The strain tensors $u_{ij} ( \vec{r} )$ associated to the deformation of the graphene layer depend both
on the in-plane displacements and heights in Monge's representation:
\begin{eqnarray}
u_{11} &= &\frac{\partial u_x}{\partial x} + \frac{1}{2} \left(
\frac{\partial
    h}{\partial x} \right)^2   \label{strain}
 \\
u_{22} &= &\frac{\partial u_y}{\partial y} + \frac{1}{2}\left(
\frac{\partial
    h}{\partial y} \right)^2 \nonumber \\
u_{12} &= & \frac{1}{2} \left( \frac{\partial u_x}{\partial y} +
    \frac{\partial u_y}{\partial x} \right) + \frac{1}{2}\frac{\partial h}{\partial x}
    \frac{\partial h}{\partial y}   \nonumber
\end{eqnarray}
In a practical calculation it is convenient to use an integration domain that remains fixed for every moir\'e period. 
For this purpose we use rescaled coordinates to operate in the coordination 
vector $\vec{d}$ defined in the unit cell of graphene.
Using the chain rule to relate the reduced vector $\vec{d}$ in graphene's unit cell and the 
real-space $\vec{r}$ coordinates for zero twist angle and assuming variable lattice constant mismatch $\varepsilon$ we have
\begin{eqnarray}
\nabla_{\vec{r}} 
=   \varepsilon   \nabla_{ \vec{d} }.
\end{eqnarray}
When we neglect the contributions from the height variation the elastic energy can be written as
\begin{eqnarray}
{E}_{\rm elastic} =  \frac{\varepsilon^2}{  A_{M}} \int_{A_M} d^2  \vec{r} \,\,  S_{el} (\vec{d}(\vec{r}), \vec{u})
\label{elasticityeq}
\end{eqnarray}
where $S_{el}$ represents the integrand of Eq. (\ref{elasticityeq}) 
in rescaled coordinates $\vec{d}$. This form shows more explicitly a $\varepsilon^2$ weakening of the elastic energy
as the lattice constant mismatch becomes smaller.

\subsection{Parametrization of the potential energy}
Likewise it is convenient to use the parametrization of the potential energy in the 
coordination vector $\vec{d}(\vec{r})$ and the interlayer separation height. 
The potential energy term has been parametrized from EXX+RPA calculations
binding energy curves for different stacking configurations \cite{supp_sachs}
as a starting point to extract the potential energy curves needed for the formulation of the 
Frenkel-Kontorova (FK) model for this two-dimensional bipartite lattice.
We can neglect the van der Waals tail corrections from the bulk that bring the equilibrium distances
closer because their influence in distinguishing different stacking energies are small.
We make use of the property that the energy landscape for a fixed $z$-axis separation is given by 
a simple expansion in the first shell of G-vectors in Fourier space \cite{supp_moirebandtheory}
to represent the energy map with three parameters. 
As noted previously the simplest approximation for 
a scalar field that varies smoothly in real space
with the triangular lattice symmetry is given by
\begin{eqnarray}
\Phi\left( \vec{r} \right) &=& \sum_{\vec{G}} C_{\vec{G}} \exp \left( - i \vec{G} \cdot \vec{r} \right)  \simeq  C_0 +  f_1(\vec{r},C_1,\varphi_1)      \nonumber
\end{eqnarray}
where the constants $C_{\vec{G}}$ are complex numbers.
Thanks to this simple form it is possible to parametrize the whole energy landscape
from the values of the potentials at three inequivalent stacking configurations, 
for example the three symmetric stacking configurations AA, AB and BA.
Its explicit expression
\begin{eqnarray}
\label{potkernel}
\Phi(x,y, C_0, C_1, \varphi)  &=& C_0 + 2 C_1 \cos (\varphi - G_1 y)    \label{potkernel}
\\  & +& 4 C_1 \cos (G_1 y / 2 + \varphi ) \cos(\sqrt{3} G_1 x / 2) \nonumber 
\end{eqnarray}
repeats with the periodicity of a triangular lattice. 
The scalar function at the three distinct symmetry points in units of graphene's
lattice constant 
\begin{eqnarray}
A &=& \Phi(0,0)   
= C_0 + 6 C_1 \cos \varphi     \\
B &=& \Phi(0, \frac{1}{\sqrt{3}}) \\
&=& C_0 + 2 C_1 \cos \left( \varphi - 4 \pi / 3 \right)    + 4 C_1 \cos \left( 2 \pi /3 + \varphi \right)     \nonumber  \\
C &=& \Phi(0, \frac{2}{\sqrt{3}}) \\
&=& C_0 + 2 C_1 \cos \left( \varphi - 8 \pi/3 \right) + 4C_1 \cos \left( 4 \pi/3 + \varphi \right).    \nonumber
\end{eqnarray}
These equations lead to the explicit values of the parameters
\begin{eqnarray}
\varphi &=& \arctan \left[ - \frac{\sqrt{3}}{2(D + 1/2)} \right] \\
C_1 &=& \frac{C - B}{6 \sqrt{3} \sin \varphi} \\
C_0 &=& - 6C_1 \cos \varphi + A
\end{eqnarray}
where we have used the relation $D = (A - B) / (B - C)$.
From the layer separation $z$-dependence of these three coefficients
$C_0(z)$, $C_1(z)$ and $\varphi(z)$ we can obtain the complete potential landscape 
$U(x,y,z)$ 
that we need for our model. 
We note that the $C_0(z) = (A(z) + B(z) + C(z))/3$ term is the average value of $\Phi(x,y)$ in the periodic domain
for every value of $z$ and that the remaining $C_1(z)$ and $\varphi(z)$ terms accounts for the landscape
of the energy in the first harmonic approximation, which is often an accurate approximation for functions varying
smoothly with the moir\'e pattern \cite{supp_moirebandtheory}.
The difference between this average and the minimum $U_{\rm dif} = U_{\rm av} - U_{\rm min}$ gives a measure of 
the in-plane forces associated to the energy gradient in a Frenkel-Kontorova problem \cite{supp_frenkelkontorovabook}. 
The numerical values for $A(z)$, $B(z)$ and $C(z)$ for the binding energy curves 
as a function of separation distance $z$ can be obtained from the calculations provided
in Ref. \cite{supp_sachs}.
They can be interpolated numerically or alternatively we can use analytic fitting expressions
similar to that in Ref. \cite{supp_gould} used in the G/G case.
We define the auxiliary functions
\begin{eqnarray}
M(x) &=& -M_0 ( 1 + \tau x ) \exp( -\tau  x ) \\
T(x) &=&  T_0 / (x^4 + T_1 )  \\
W(x)   &=&  (1 + \exp( -16 (x - 4 ) ))^{-1}
\end{eqnarray}
with the parameters $M_0 = 0.06975$, $\tau = 7$, $D_0 = 3.46$, $T_0 = - 10.44 $, $ T_1 = -58.87$
to define the fitting function for the average value of $C_0(z)$ for all the stacking configurations through
\begin{eqnarray}
C_0(z) &=& M\left(  z/ D_0 - 1 \right)  \\
&+& \left( T(   z ) - M\left(  z / D_0 - 1 \right)  \right)  W(  z)
\end{eqnarray}
where $z$ is given in angstroms.
We used a rather simple model for $W(x)$ which is fairly accurate but
can still be improved through additional parameters to better capture the behavior away from the equilibrium point.
The $z$-dependence of the $C_1(z)$ term is easily captured through an exponentially decaying form
\begin{eqnarray}
C_1(z) = a \exp(- b (z/a_0 - z_0)) 
\end{eqnarray}
where $a = 2.226$, $b= -3.295$ and $z_0 = 1.295$, $a_0 = 2.46 \, \AA$ is the lattice constant
of graphene.
The $\varphi = -50.4^{\circ}$ term shows a weak linear dependence with respect to $z$ 
so we use a constant value.
When necessary, the long-ranged van der Waals tails originating from the bulk 
BN layers can be added through 
\begin{eqnarray}
T_{\rm tail}(z)  = \sum_{n} T(z + n c)
\end{eqnarray}
where $c$ is the separation lattice constant between the layers
and whose sum saturates quickly.
However, this correction term has a small influence for the differences in energy
for different stacking arrangements and we neglect this term.
The energy landscape plots for a fixed separation distance $z_0 = 3.4\, \AA$
presented in Fig. \ref{landscape}
\begin{figure}[htbp]
\begin{center}
\includegraphics[width=8.4cm,angle=0]{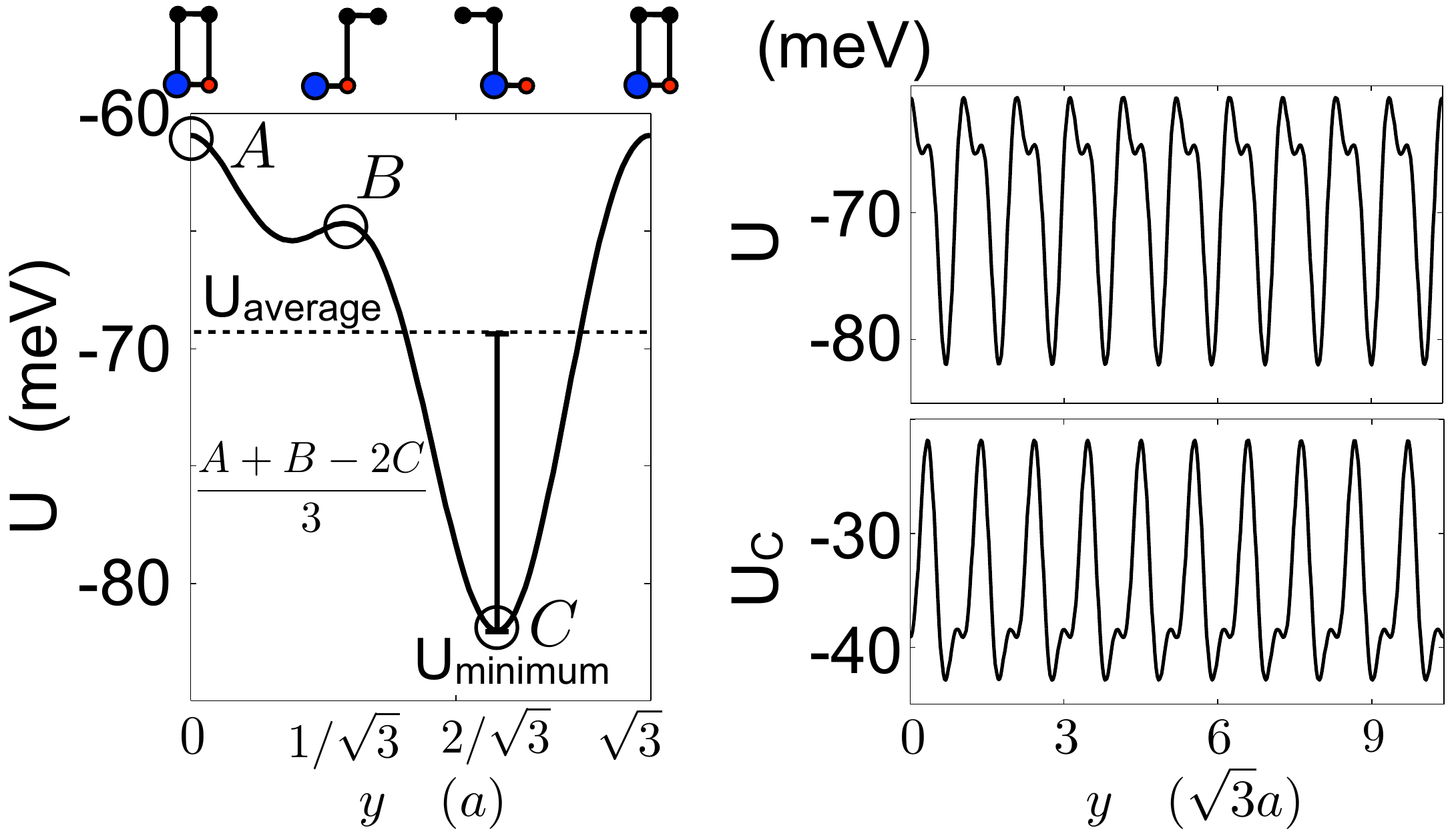}
\end{center}
\caption{ {\em Left Panel:} The total energy per unit cell area as a function of sliding in the $y$ axis
for $x=0$ shows a minimum when one of the carbon atoms sits 
in the middle of the hexagon and another sits on top of boron. 
{\em Right top panel:}
Potential energy of graphene's carbon atoms per unit cell area.
{\em Right bottom panel:} potential energy experienced by the individual carbon atom per unit cell area
obtained assuming additivity of the energies. 
}
\label{landscape}
\end{figure}
allows to estimate the average 
in-plane traction force being applied on the two inequivalent carbon atoms in the unit cell. 
Even though the LDA binding energies are substantially smaller than in an EXX+RPA calculation, 
we find that this in-plane energy map obtained through parametrization in Fig. 3 of the main text 
is closely similar to the LDA energy map obtained in Ref. \cite{supp_moirebandtheory}, 
whose agreement is attributable to the dominance of short-range character of the interactions
near equilibrium distances that is captured reasonably well by the LDA approximation \cite{supp_gould}.

From this potential landscape per two carbon atom unit cell we can infer the 
potential experienced by the individual carbon atoms that can be useful for lattice force-field calculations
where the higher energy optical modes are treated explicitly. 
This is done assuming that the total potential energy consists of the sum of the potentials 
experienced by each carbon atom which is separated by a distance $\tau = a/\sqrt{3}$
\begin{eqnarray}
U(x,y) = U_{C}(x,y) + U_{ C}(x,y+\tau).
\end{eqnarray}
Solving the above equation we get 
\begin{eqnarray}
U_{ C}(x,y) = \Phi(x,y,C_0',C_1,\varphi'_1)  
\end{eqnarray}
where $C_0' = C_0/2$ and $\varphi_1' = \varphi_1 - \pi/3$. 
Likewise if the long-range van der Waals tails are used they would need to be reduced to one half of its value.
In Fig. \ref{landscape} we show the potential energy repeated over several periods 
as well as the energy landscape seen by each carbon atom, derived assuming additivity in the total 
potential energy.

\section{Relaxation of the graphene and substrate atoms in the G/BN heterojunction}
The relaxed geometries can be readily obtained minimizing the elastic and potential energy functionals
using the trial functions that we introduced earlier. 
We distinguish two different scenarios in our elasticity problem.
In the first case we solve for the elastostatic solutions where we relax the atoms of graphene 
subject to a periodic moir\'e potential of a rigid substrate. A second scenario allows the coupled
relaxation of the BN substrate atoms. 
For simplicity we consider only the zero twist angle case and variable lattice constant mismatch $\varepsilon$.

\subsection{Graphene relaxation only model}
\begin{figure}[htbp]
\begin{center}
\includegraphics[width=8.2cm,angle=0]{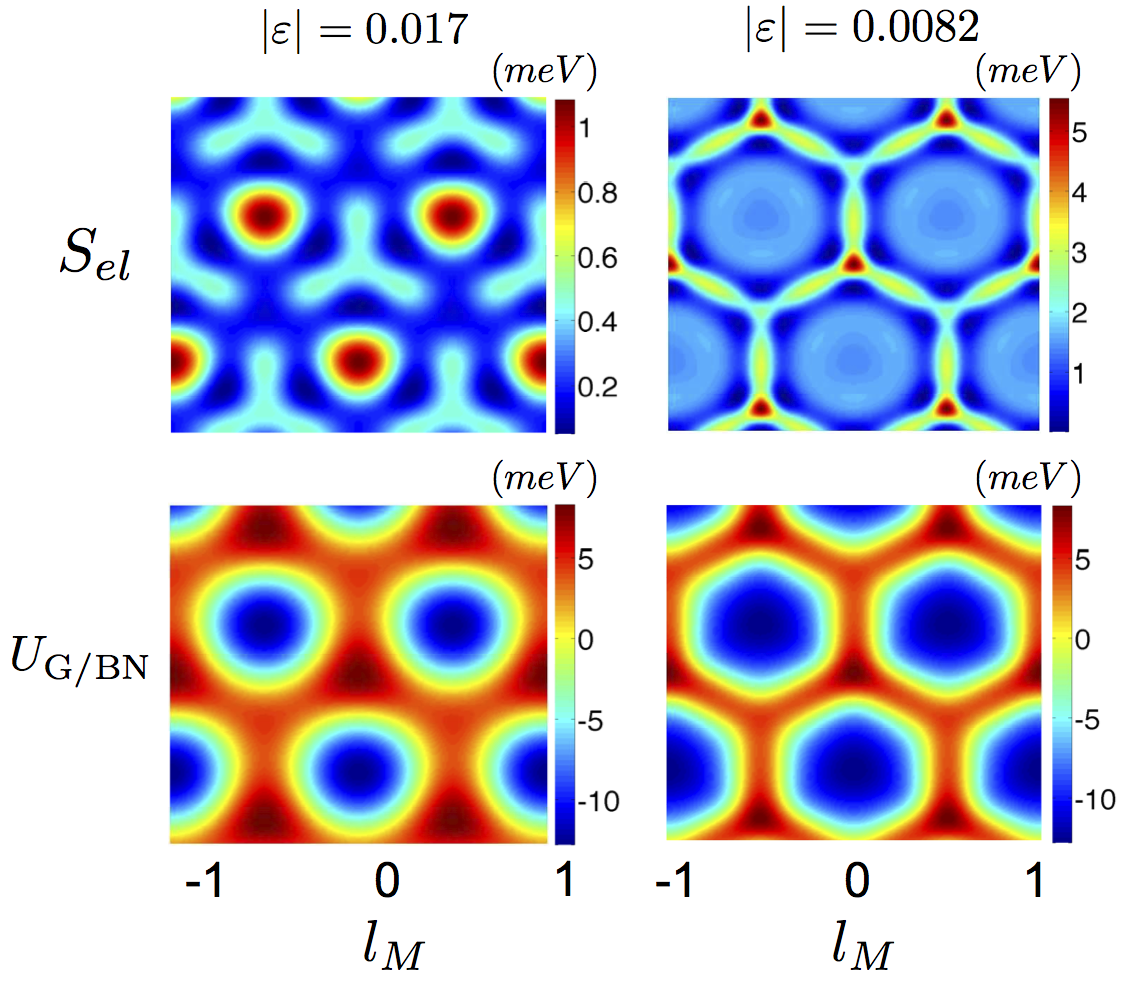}
\end{center}
\caption{ 
Map of local elastic and potential energies per unit cell area (see Eqs. \ref{elasticityeq}, \ref{potkernel}) corresponding 
to small and large strains using \ constant $h$ model, see also \cite{supp_apstalk}, corresponding to 
lattice constant differences of of $\left| \varepsilon \right| = 0.017, 0.0082$ repectively.
The large strain configuration we represent here is just before the poitnt of steep transition
as shown in Fig. \ref{elastostatic_g}, which happens for longer moir\'e periods than when
the $z$-axis relaxation is allowed.
}
\label{elastpotenergymap}
\end{figure}

The resolution of the elastostatic problem of graphene subject to a superlattice potential
requires the minimization of the total energy functional
\begin{eqnarray}
{E}_{\rm total} = {E}_{\rm elastic} + { E}_{\rm potential}
\end{eqnarray}
where ${E}_{\rm elastic}$ is given in Eq. (\ref{elasticityeq}) and the potential energy is given by
the integral in the moire supercell of area $A_M$ of the potential energy kernel 
$U (\vec{r},\vec{d},h)$ given in Eq. (\ref{potkernel})
\begin{eqnarray}
{E}_{\rm potential} = \frac{1}{A_M} \int_{A_M} d \vec{r} \,\, U(\vec{r},\vec{d},h).
\end{eqnarray}
The stacking coordination vector is modeled as 
\begin{eqnarray}
\vec{d}_G(\vec{r}) = \vec{d}_0 (\vec{r}) + \vec{u}_{G}(\vec{r})
\end{eqnarray}
assuming that the substrate produces a rigid periodic potential pattern.
We use the gradients of the scalar field in Eq. (\ref{vectorfield})
to model the displacement vectors $\vec{u}_G$. 
The local elastic and potential energy maps corresponding to 
the small and large strain limits are represented in Fig. \ref{elastpotenergymap}.

\subsection{Coupled relaxation of the BN lattice}

\begin{figure}[htbp]
\begin{center}
\includegraphics[width=8.2cm,angle=0]{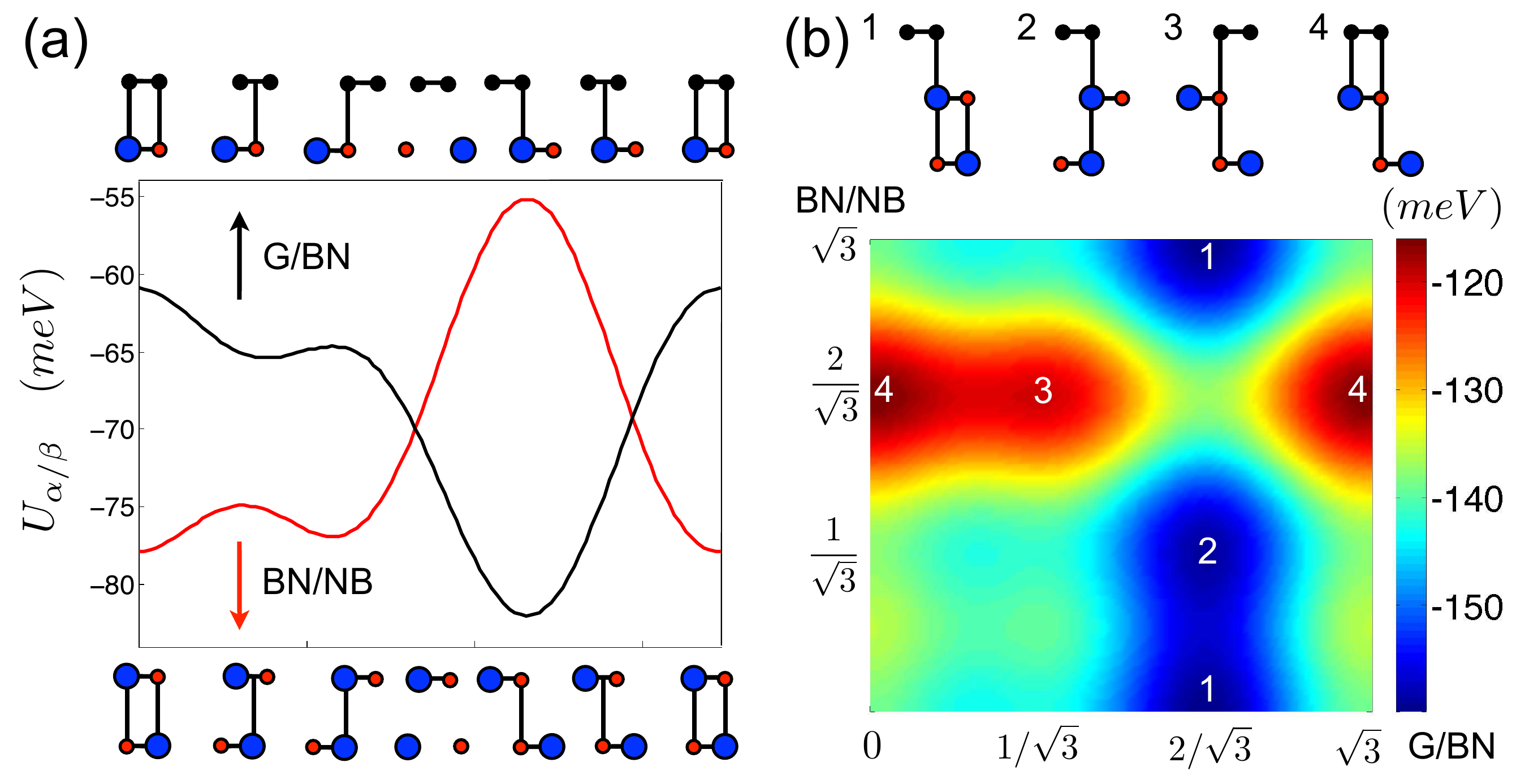}
\end{center}
\caption{ 
In (a) we show the total energy per unit cell area for sliding along the vertical $y$-axis
for different stacking configurations for G/BN within RPA and BN/NB heterojunctions
within LDA near the equilibrium interlayer separation for fixed $c = 3.4 \, \AA$.
The energy curves were obtained 
using the information at three
different symmetric stacking configurations for AA, AB and BA for interlayer sliding vectors
$\tau_{AA} = (0,0)$, $\tau_{AB} = (0, a / \sqrt{3})$ and $\tau_{BA} = (0, 2 a / \sqrt{3})$.
The energy minimum for G/BN stacking happens at $\tau_{BA}$ whereas 
for BN/NB the energies are smallest near $\tau_{AA}$ and $\tau_{AB}$.
In (b) we show the stacking configurations for G/BN and BN/NB and minimize the total energy
which shows that deformation of the topmost BN layer is easier when it preserves the BA stacking 
of the G/BN hererojunction. The minimum energy configurations are indicated with labels 1, 2 whereas
the maximum energy ones are labeled with 3, 4.
}
\label{energylandscapemaps}
\end{figure}

Here we explore the influence in the elastic of energy of graphene when the 
BN atoms of the topmost layer in the substrate are allowed to relax in 
response to the stacking rearrangement of the graphene sheet.
The coupled motion of the substrate atoms
contribute in decreasing the total elastic energy of the graphene BN heterojunction 
because a smaller displacement in the graphene sheet is needed than if the substrate remains rigid.
For solving the coupled G/BN elasticity problem we will assume that 
the topmost BN sheet is subject to a potential stemming from the graphene sheet itself
and the BN layers underneath, assuming that the BN atoms below the topmost layer remain fixed.
The potential energy for fixed interlayer separation of $c = 3.4\, \AA$ for G/BN and BN/NB
along the $y$ direction of stacking arrangement vector is shown in Fig. \ref{energylandscapemaps}. 
The interaction potential between the two topmost BN layers are defined by 
$C_1 = -2.47$~meV and $\varphi_1 = -57.75^{\circ}$ through the funciton 
$U_{\rm BN/NB}(\vec{r}; \vec{d}_{BN}) = f_{1}(\vec{d}_{BN}, C_1,\varphi_1)$.

Even though the binding energies and forces predicted by the LDA
typically underestimate the values obtained from higher level RPA calculations \cite{supp_jelliumrpa,supp_hbnacfd,supp_graphitebinding,supp_bjorkman}
we assume that the energy landscape for different strackings 
We used LDA energies for BN/NB coupling as a function of sliding, 
assuming that their sliding energy maps are comparable to EXX+RPA
as we found for the G/BN case.
The total energy of G/BN/NB where both sheets are allowed to relax is given by 
 the sum of the elastic and potential energies of graphene and the topmost BN sheet.
The total potential energy term can be obtained 
from the interaction energies between the neighboring layers through
\begin{eqnarray}
{ E}_{\rm potential} &=& E_{\rm potential, \, G/BN}  +  E_{\rm potential, \, BN/NB},   
\label{poteng}
\end{eqnarray}
and can be calculated from the parametrized potential energies evaluating the integrals in the moir\'e supercell
\begin{eqnarray}
{ E}_{\rm potential, \, G/BN} &=&   \frac{1}{A_M} \int_{A_M}  d \vec{r}  \,\, \,  U_{\rm G/BN}(\vec{r} \,; \, \vec{d}_G , h_G),   \\
{ E}_{\rm potential, \, BN/NB} &=&   \frac{1}{A_M} \int_{A_M}  d \vec{r}  \,\, \,  U_{\rm BN/NB}(\vec{r} \,; \, \vec{d}_{BN} ),
\end{eqnarray}
where the kernels are functionals of the local stacking coordination functions $\vec{d}_G$ and $\vec{d}_{BN}$
that depend on the displacements relative to the neighboring layers
\begin{eqnarray}
\vec{d}_{G}(\vec{r})  &=& \vec{d}_0(\vec{r}) + \vec{u}_{G}(\vec{r}) - \vec{u}_{BN}(\vec{r}), \\
\vec{d}_{BN}(\vec{r})  &=&  \vec{u}_{BN}(\vec{r}).  \\   \nonumber
\end{eqnarray}
We used explicit labels  G/BN and BN/NB to distinguish the interaction potentials.
For the graphene sheet the only relevant reference frame is the topmost BN layer 
whereas the latter interacts both with the graphene sheet and the BN layers underneath
whose coordinates are assumed to remain fixed.

\begin{figure}[htbp]
\begin{center}
\includegraphics[width=8cm,angle=0]{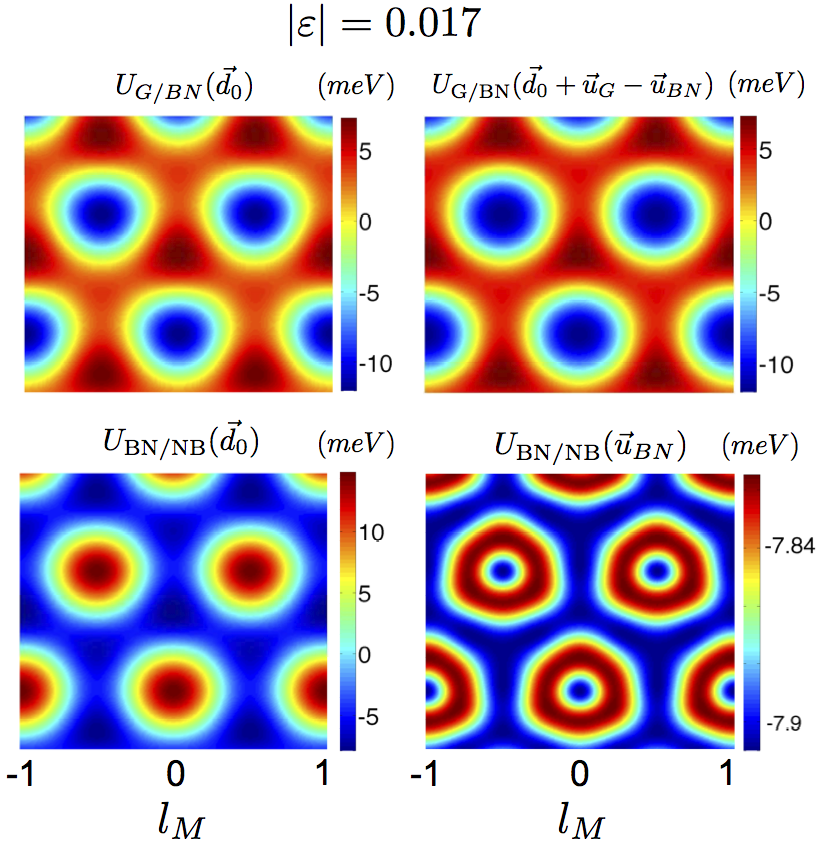}
\end{center}
\caption{ Potential energy maps $U_{\rm G/BN}$ and $U_{\rm BN/NB}$ in real space. 
The left column represents the interaction potentials of rigid graphene and BN sheets.
The right column shows the potential energy maps for  $U_{\rm G/BN}$ and $U_{\rm BN/NB}$ 
corresponding to the relaxed geometry configurations determined by 
the strains in the graphene and boron nitride sheet given by $\vec{u}_{G}$ and $\vec{u}_{BN}$ respectively.
}
\label{elastostatic_gbn}
\end{figure}

\section{Strains in relaxed ground-states}
\begin{figure}[htbp]
\begin{center}
\includegraphics[width=8.2cm,angle=0]{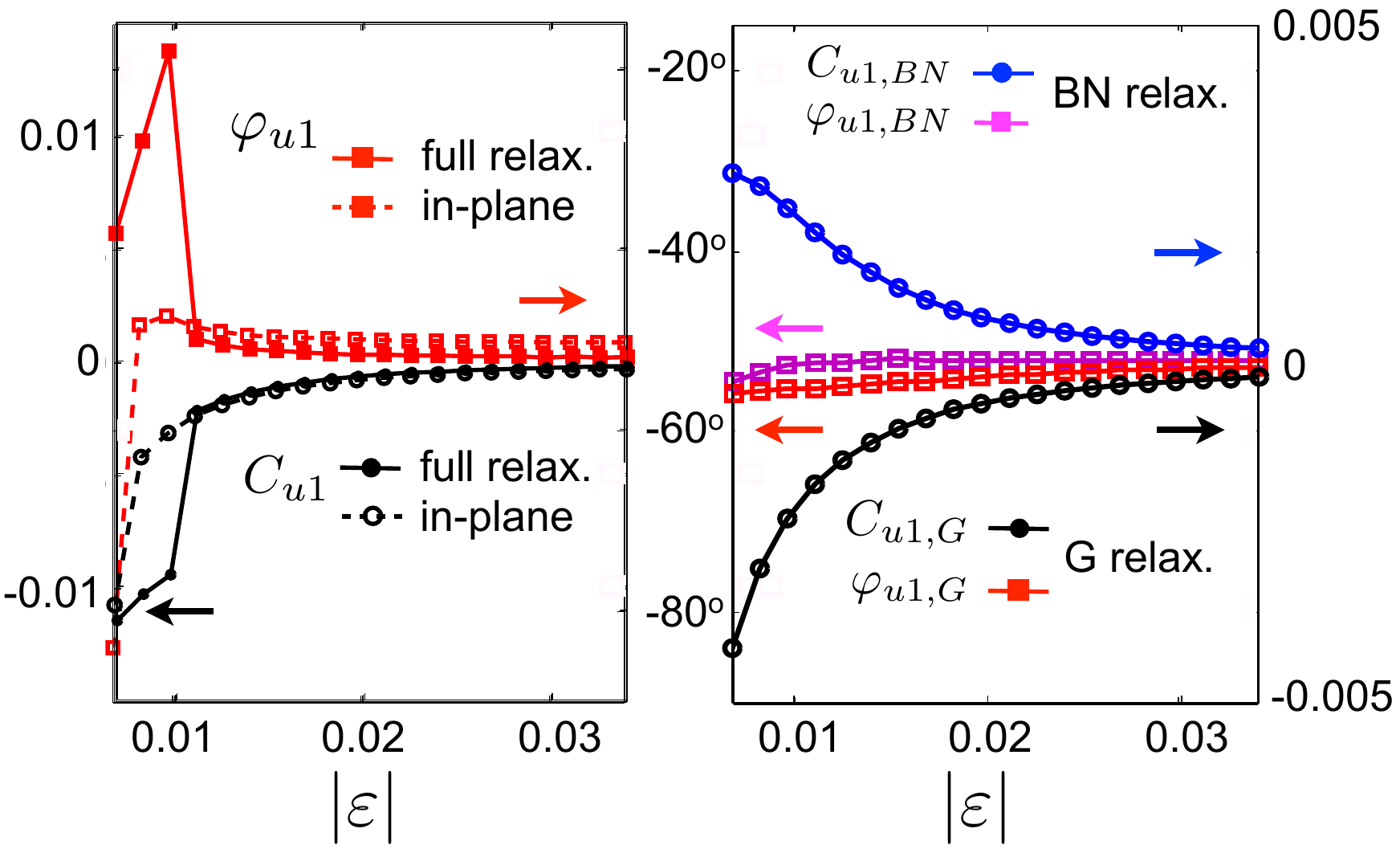}
\end{center}
\caption{ 
{\em Left panel:}
Elastostatic solutions for the strains in the graphene sheet relaxation only 
that is subject to the potential of a rigid BN substrate. 
For smaller $\left| \varepsilon \right|$ 
the potential energy dominates and the deformation becomes larger. 
The increase of the deformation is steady until it reaches a tipping point where the solutions become unstable. 
Comparison of in-plane relaxation only and that allowing out-of-plane relaxation shows that the both approximations
give similar in-plane displacements but allowing the full relaxation makes the transition easier. 
{\em Right panel:}
Elastostatic solutions of the coupled graphene and topmost BN layer subject to the potentials of a rigid BN layer potential underneath.
We notice that the magnitude of the in-plane deformation of the graphene and BN sheets are comparable 
to the strains in the BN sheet as the latter can relax along the easy sliding axis directions.
}
\label{elastostatic_g}
\end{figure}

As noted earlier, the strained geometries can be characterized by the magnitude and phases
that define the scalar fields in Eq. (\ref{scalarfield}) and vector fields in Eq. (\ref{vectorfield})
within a restricted variational space that preserves the triangular moir\'e periodicity dictated by 
the lattice constant mismatch $\varepsilon$ and twist angle $\theta$.
The solutions for the strains in the graphene layer can be largely characterized by two parameters $C_{u,1}$ and $\varphi_{u, 1}$
whereas the height profiles require three $C_{h,0}$, $C_{h,1}$ and $\varphi_{h,1}$ for the additional average 
interlayer separation. For the topmost BN sheet we only consider in-plane strains assuming that its separation from the
additional BN layer underneath takes a constant average value. 
Because the relative magnitudes of the elastic and potential energies scale with the lattice constant mismatch $\varepsilon$
a correction in the potential profiles or the average value of the elasticity constants would have an overall
effect of shifting the solutions in the abscissa. 
Within our approximation, the solutions are completely characterized by the $\varepsilon$-dependent values 
of the parameters that define the scalar field. 
In Fig. \ref{elastostatic_g} we show the values of the relaxed solution parameters where only the graphene
sheet is allowed to relax both in-plane and out of plane.
We also show a comparison with the solutions where only in-plane relaxation is permitted and we fix
the interlayer separation to a constant value. 

When we consider the coupled graphene and BN layer relaxation we notice an interesting behavior
where the largest strain magnitudes are for the topmost BN sheet rather than graphene itself.
The results for the relaxed strains of the coupled G/BN/NB heterojunction 
where both graphene and the topomost BN layer are relaxed is shown in Fig. \ref{elastostatic_gbn}.
This is possible thanks to a special total energy landscape with easy sliding path in BN \cite{supp_marom}.
The BN sheets in the crystal substrate follows an AA$^{\prime}$ stacking order and for this stacking configuration
they have a minimum energy sliding path when going from AA$^{\prime}$ to AB$^{\prime}$ with a small barrier of about  $\sim$3 meV
and even smaller total energy differences of about  $\sim$1 meV within the LDA. A more elaborate GGA + vdW
functional calculation \cite{supp_marom} predicted similar barrier magnitudes but with the minimum of energy
happening for the AB$^{\prime}$ stacking configuration with a total energy lower by $\sim$1.5 meV.
These minute differences are unimportant for the solutions we discuss.

\section{Band gaps and the Fourier components of the strained Hamiltonian}
\begin{figure}[htbp]
\begin{center}
\includegraphics[width=8.8cm,angle=0]{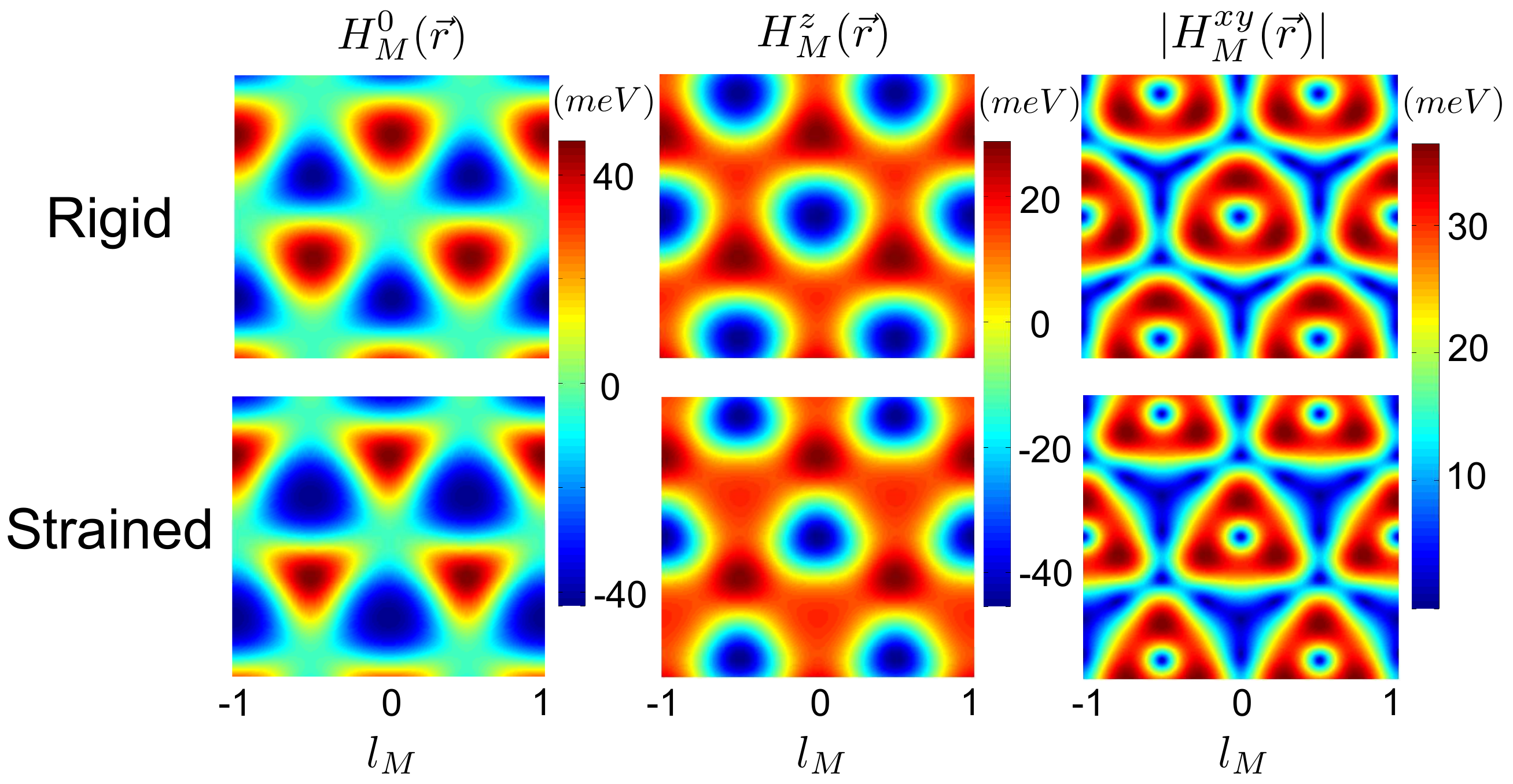}
\end{center}
\caption{ 
Real space representation of the pseudospin Hamiltonian
for unrelaxed (top row) and relaxed (bottom row) geometries near zero twist angle.
The $H^0_{M}$ term accounts for the site potential fluctuations normally seen in scanning probe
studies, the $H^z_{M}$ term is the mass term dictating the local band gap in real space, 
and $ H^{xy}_{M} $ reflects the anisotropic strains.
}
\label{pseudospins}
\end{figure}

\begin{figure}[htbp]
\begin{center}
\includegraphics[width=7.4cm,angle=0]{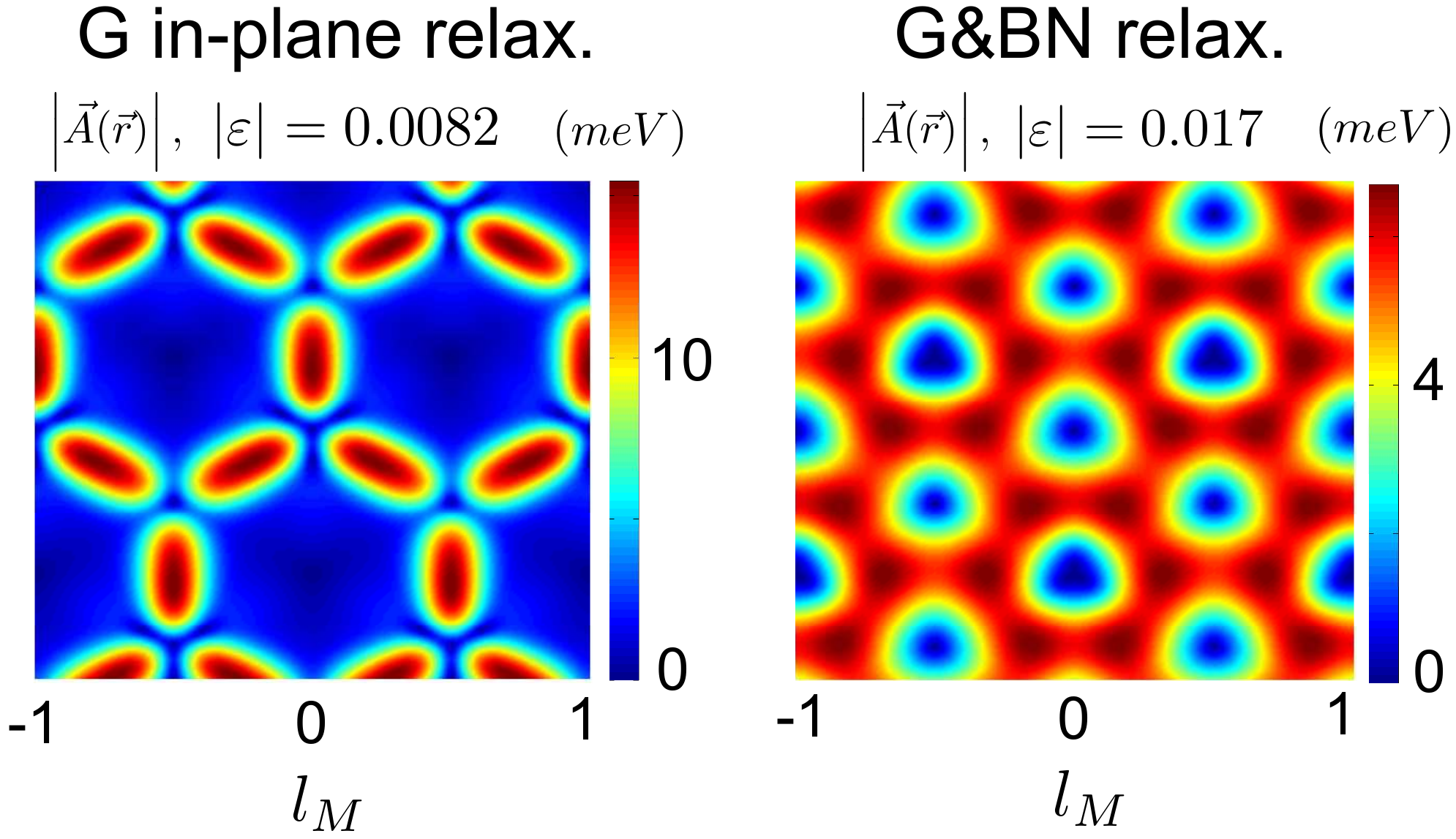}
\end{center}
\caption{ 
Real space map of the pseudomagnetic field magnitude $\left| \vec{A}(\vec{r}) \right|$
for relaxed solutions. The left panel represents graphene relaxation only allowing
only in-plane strains with a long moir\'e period corresponding to $\left| \varepsilon \right|  = 0.0082$.
On the right hand we present the magnitude of the vector potential due to the relaxation of the graphene sheet
when both graphene and BN sheets are simultaneously allowed to relax near zero twist angle moir\'e period. 
}
\label{pseudomagneticcontour}
\end{figure}

\begin{figure}
\includegraphics[width=\columnwidth]{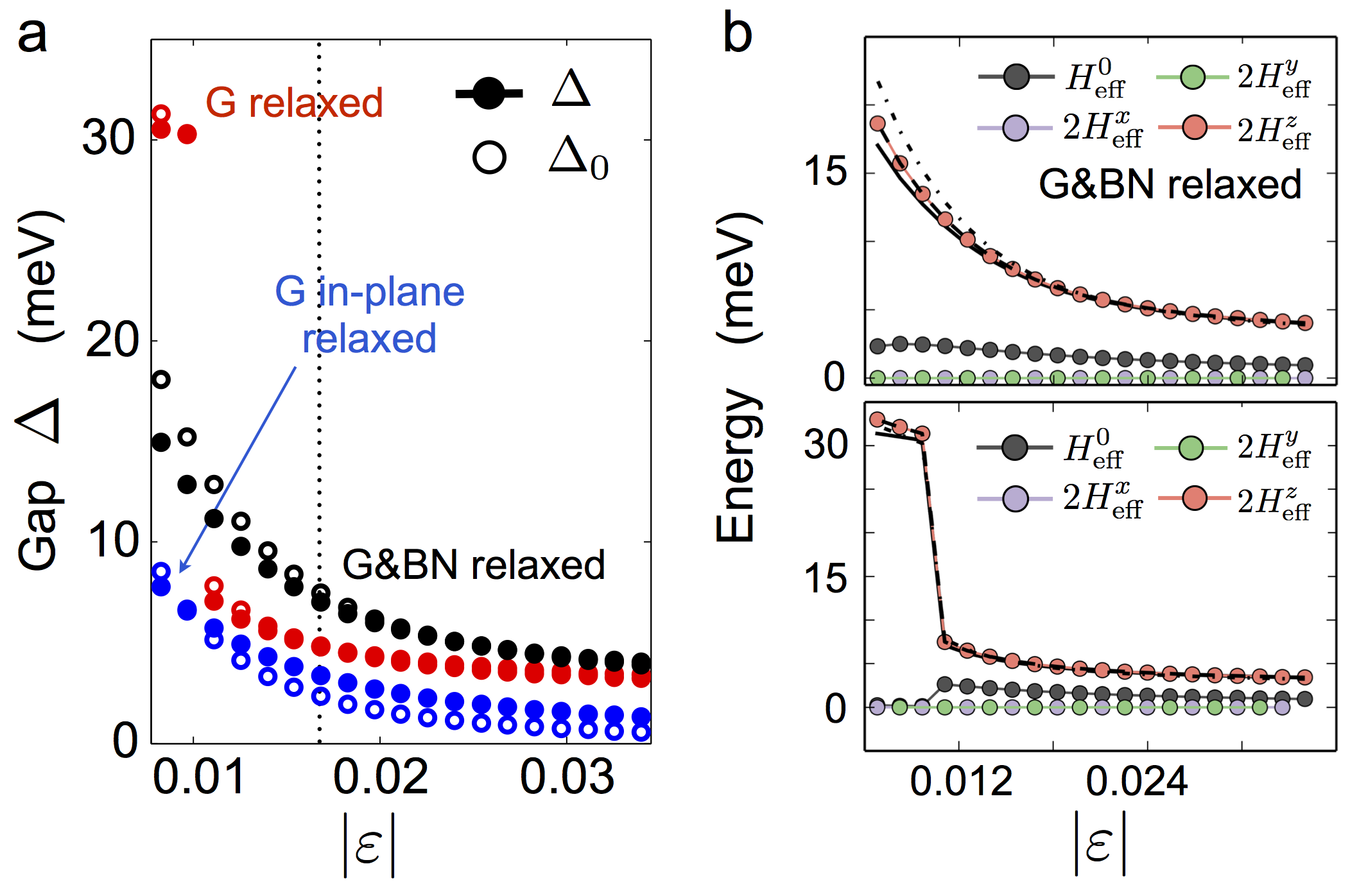}  
\caption{
Breakdown of different contributions to the single-particle band gap.
{\em Left panel:} Comparison of the band gap $\Delta$ represented with connected 
filled circles and the non-zero average contribution $\Delta_0$ represented with 
empty circles for graphene relaxed, graphene and boron nitride relaxed, and restricted in-plane only relaxation
of graphene. 
We notice that the presence of out-of-plane relaxation prevents the complete cancellation
of the average mass in the presence of small in-plane strains.
{\em Top right panel:}
Perturbation theory analysis of the gap in the configuration where graphene relaxes due to in-plane strains. The non-perturbative gap is shown in a black solid line, which is closely approximated by the 2nd order perturbation theory (dashed black line). 
The dash-dotted line is the gap due to the average mass.
The decomposition of the Hamiltonian into $H^{0}_{\rm eff}$ (grey circles) and 
$H^{\alpha}_{\rm eff}$ for $\alpha = x,y,z$ (blue, green, and red circles, respectively) 
indicates that the primary source of the gap is $H^{z}_{\rm eff}$.  
{\em Bottom right panel:} 
Perturbation theory analysis of the gap in the configuration where graphene and boron nitride both relax due to in-plane strains. 
The non-perturbative gap is shown in a black solid line, which is closely approximated by the 2nd order perturbation theory (dashed black line). 
The dash-dotted line is the $\vec{G}_0 = 0$ contribution to the gap. 
The decomposition of the Hamiltonian into $H^{0}_{\rm eff}$ (grey circles) and $H^{\alpha}_{\rm eff}$ for 
$\alpha = x,y,z$ (blue, green, and red circles, respectively) indicates that the primary source of the gap is $H^{z}_{\rm eff}$.}
\label{gapcomponents}
\end{figure}

The Hamiltonian of graphene is modified in a G/BN heterojunction by moir\'e patterns \cite{supp_moirebandtheory}
that can be described in a transparent manner when represented in a pseudospin basis.
The different contributions consist of a site potential $H^{0}_M$, the mass or sublattice staggering potential $H^z_{M}$
and an in-plane pseudospin inter-sublattice coherence term $H^{xy}_{M} = H^{AB}_{M}$ as shown in Fig. \ref{pseudospins}.
The latter is closely related with a pseudomagnetic field derived from the straining of the graphene
sheet represented in Fig. \ref{pseudomagneticcontour}.
As mentioned in the main text and explained in more detail in Ref. \cite{supp_moirebandtheory}
it is possible to obtain the full band structure from the Fourier components represented 
in the moir\'e reciprocal lattice vectors $\vec{G}$.
The term that most directly influences the band gap comes from $H^z_{M}$, in particular the 
$\vec{G}_0 = (0,0)$ contribution which is the average mass in the a moir\'e supercell.
This term normally vanishes to zero in a rigid crystal \cite{supp_moirebandtheory} 
but here we showed that they generally average to a nonzero value in the presence of in-plane strains.
Farther shell contributions in $\vec{G}$ do also contribute to the band gap through higher order corrections.
The contributions from the first shell in G-vectors through second
order perturbation theory play the most relevant role.
In Fig.~\ref{gapcomponents} we show a comparison of the nonzero average mass term and the total
band gap as a function of moir\'e period.
It is noteworthy that when out of plane relaxations are absent the values of the band gaps 
are generally smaller and the relative cancellation between the
average mass $\Delta_0$ and the second order contributions from the first shell are more substantial.
The pseudomagnetic field term from $H^{AB}_{M}$, due to the anisotropic strains generated by the coupling with the 
BN substrate, plays a minor role in configuring the band gap at the primary Dirac point as will be made clearer
in the perturbative analysis.
\begin{figure}[htbp]
\begin{center}
\includegraphics[width=7.4cm,angle=0]{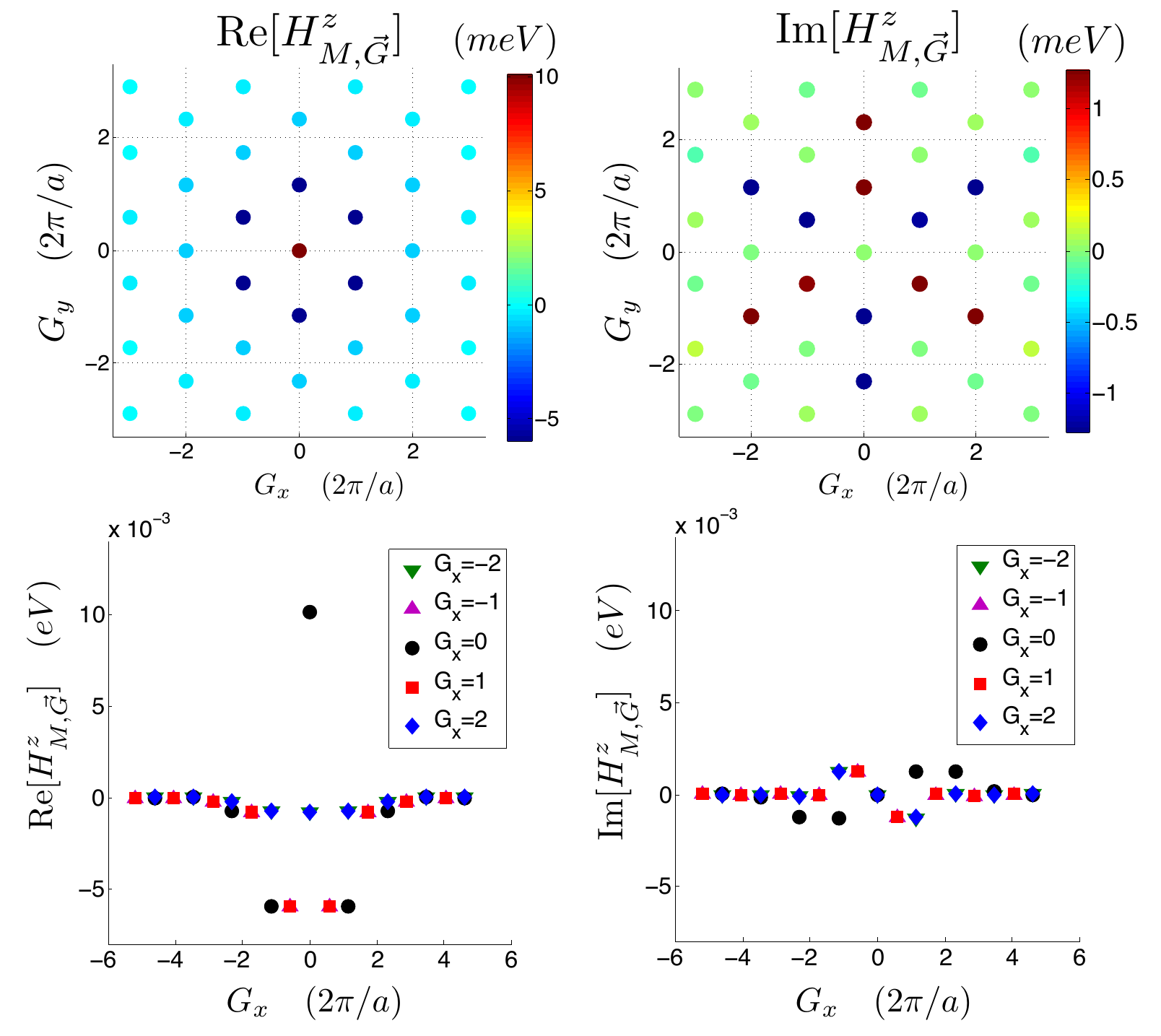}
\end{center}
\caption{ 
A representation of the real and imaginary parts of the Fourier components of the $H^{z}_{M}$ local 
mass term distribution in real space of the Hamiltonian for substantially strained configurations for in-plane only
relaxation of the graphene sheet. 
The band gap is determined mainly by the average mass term from the $\vec{G}=(0,0)$ contribution and modified 
by the first hexagonal shell in $\vec{G}$ vectors contributing to second order in perturbation theory.
}
\label{hzfourier}
\end{figure}
The other two components of the Hamiltonian, potential fluctuations $H^0_{M}$ and
in-plane pseudospin terms $H^{xy}_{M}$ represented in Figs. (\ref{hzfourier}, \ref{h0xyfourier}),
also see modifications due to straining, typically acquiring contributions beyond the first shell in G-vectors
with the most important contributions ranging up to three nearest neighbor hoppings,
and their Fourier components having magnitudes in the order of $\sim$10 meV. 
\begin{figure}[htbp]
\begin{center}
\includegraphics[width=7.4cm,angle=0]{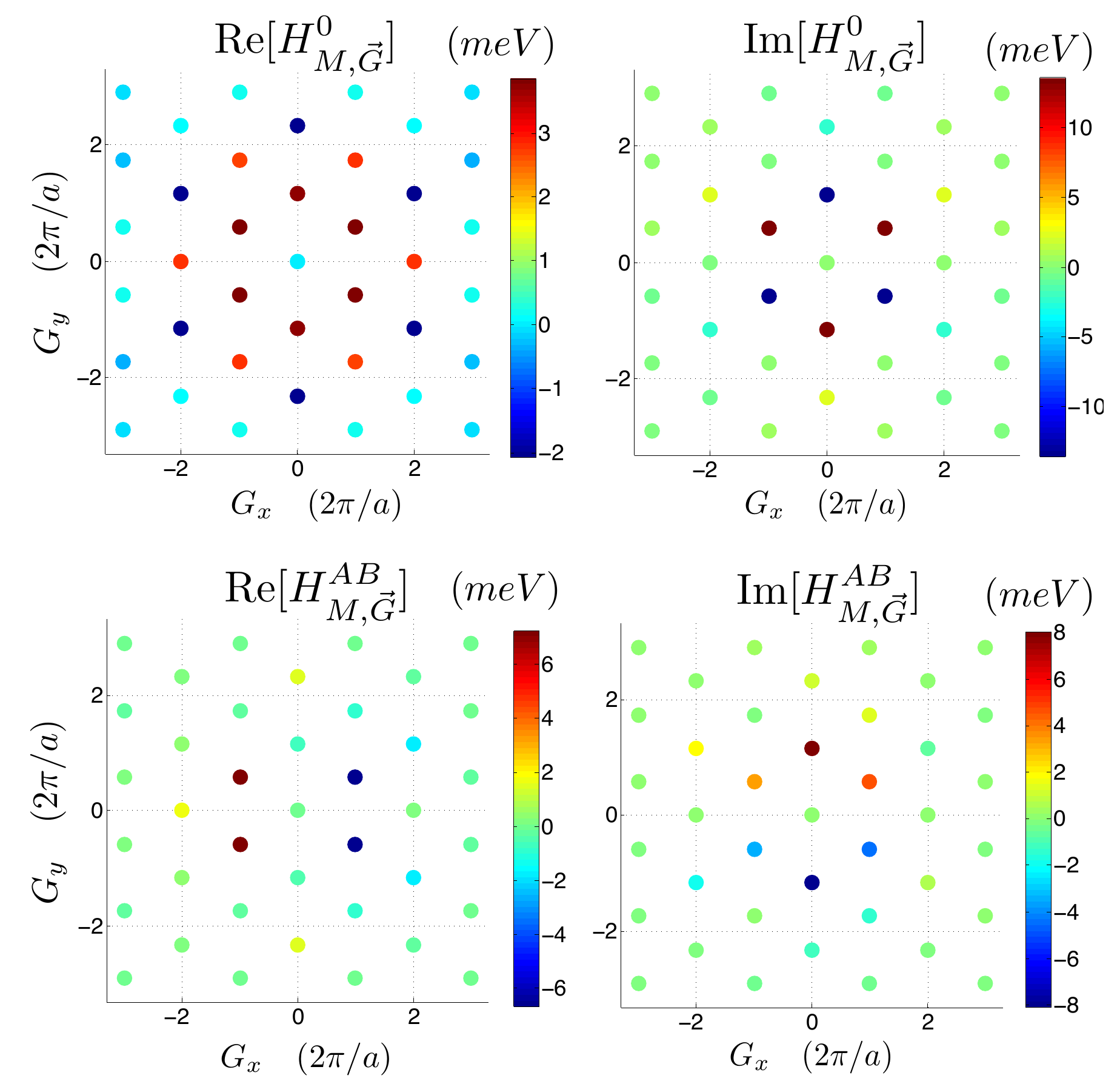}
\end{center}
\caption{ 
Fourier expansion of the Hamiltonian for the site potential fluctuations $H_0$ and the anisotropic strain 
$H_{AB}$ terms. The contributions to the band gap of these two terms are much smaller than those from 
$H_{z}$.
}
\label{h0xyfourier}
\end{figure}
The pseudomagnetic term generated by the strains in the graphene 
sheet itself can in principle have a contribution comparable to the contribution
due to the electron virtual hopping to and back from the BN sheet.
Using the expressions for pseudomagnetic fields in graphene provided in Refs. \cite{supp_suzuura,supp_vozmedianopseudo}
\begin{eqnarray}
A_x(\vec{r})  &=& g [u_{11}(\vec{r}) - u_{22}(\vec{r})]  \\
A_y(\vec{r})  &=&  - 2 g u_{12}(\vec{r})
\end{eqnarray}
and using $g  \sim 1.5/a$ a typical map of its magnitude in real space is shown 
in Fig. \ref{pseudomagneticcontour}.
We note that the pseudomagnetic vector potentials follow a moir\'e period scaling relation
given by $\left| \vec{A}(\vec{r})\right| \propto  \left( {a}/{l_M} \right) \left| \widetilde{A}(\vec{d}(\vec{r}))\right|$
when represented in rescaled coordinates of the moir\'e superlattice $\widetilde{A}(\vec{d}(\vec{r}))$, 
in turn defined by the parameters that determine the displacement vectors $\vec{u}(\vec{r})$.

\section{Second order perturbation theory}
Further insight on the contributions to the band gaps can be achieved from second order perturbation theory
from the first shell approximation. We distinguish two scenarios, one for rigid unrelaxed lattices 
and another where strains are allowed to modify the stacking coordination. 
Formally it is possible to show that for rigid unrelaxed lattices the in-plane $H^{AB}_{M, \vec{G}}$ gives a zero contribution 
to the band gap to second order in perturbation theory.
When in-plane strains are allowed, band gaps develop thanks primarily to a nonzero average mass 
and all three pseudospin components make a nonzero contribution to the gap to second order.
Among these, in our calculations the in-plane pseudospin terms contribute to the gap with a smaller magnitude than the 
Fourier expansion of the mass terms $H^{z}_{M, \vec{G}}$.

\subsection{Unrelaxed configuration}
Here we discuss the effective 2x2 Hamiltonian obtained from perturbation theory around the Dirac point. Our initial Hamiltonian is a $2N\times 2N$ matrix, where $N$ is two times the number of Moir\'e reciprocal lattice vectors in the Fourier transform. Treating the $2\times 2$ diagonal blocks as the unperturbed Hamiltonian, second order degenerate perturbation theory gives an effective Hamiltonian for the low energy states,
\begin{equation}
H_{\rm eff}=H_{M, \vec{G}=0}-\sum_{\vec{G}\ne 0}H_{M, \, \vec{G}} \, H_{G}^{-1}H_{M, \, \vec{G}}^{\dagger}\label{eqn:Heffective}
\end{equation}
where $H_{G}$ are the $2\times 2$ blocks in the Hamiltonian associated with the moir\'e vector $\vec{G}$, and 
$H_{M, \vec{G}}$ connect the $\vec{k}$ and $\vec{k}+\vec{G}$ blocks of the Hamiltonian. 
If we ignore for a moment the relaxation due to in-plane strains, the diagonal blocks are $H_{M, \vec{G}}=\hbar \upsilon\vec{G}\cdot\vec{\tau}$, 
which has an inverse $(H_{\vec{G}})^{-1}=\hbar\upsilon\vec{G}\cdot\vec{\tau}/(\hbar\upsilon\vec{G})^{2}$. 
We can decompose both the effective Hamiltonian and the $H_{M, \,\vec{G}}$ into terms proportional to Pauli matrices,
\begin{align}
H_{\rm eff}=&\sum_{\alpha=0,x,y,z}H_{\rm eff}^{\alpha}\tau^{\alpha}\label{eqn:hdefn}\\
H_{M, \vec{G}_{j}}=&\sum_{\alpha=0,x,y,z}M_{j}^{\alpha}\tau^{\alpha}\label{eqn:Mdefn}
\end{align}
Since $H_{\rm eff}$ is hermitian, the parameters $H_{\rm eff}^{\alpha}$ must be real numbers. 
However, each block $H_{M, \vec{G}_{j}}$ is not necessarily hermitian, so $M_{j}^{\alpha}$ are complex numbers. Plugging in the decomposed forms, and restricting to just the nearest shell of reciprocal lattice vectors $j=1,...,6$ (we use the index $j=0$ for $\vec{G}=0$), we get
\begin{align}
H_{\rm eff}^{0}= & \frac{4\hbar \upsilon}{(\hbar \upsilon \vec{G})^{2}}\sum_{j=1}^{3}\left[\operatorname{Re}M_{j}^{z}\left(\operatorname{Im}\vec{M}_{j}\times\vec{G}_{j}\right)\cdot\hat{z}-\right.\nonumber\\
& \qquad \left.-\operatorname{Im}M_{j}^{z}\left(\operatorname{Re}\vec{M}_{j}\times\vec{G}_{j}\right)\cdot\hat{z}\right]\label{eqn:heff0}\\
H_{\rm eff}^{x}=&-\frac{4\hbar \upsilon}{(\hbar \upsilon \vec{G})^{2}}\sum_{j=1}^{3}\vec{G}_{j,y}\operatorname{Im}\left\{M_{j}^{0}M_{j}^{z}\right\}\label{eqn:heffx}\\
H_{\rm eff}^{y}=& \frac{4\hbar \upsilon}{(\hbar \upsilon \vec{G})^{2}}\sum_{j=1}^{3}\vec{G}_{j,x}\operatorname{Im}\left\{M_{j}^{0}M_{j}^{z}\right\}\label{eqn:heffy}\\
H_{\rm eff}^{z}=& \frac{4\hbar \upsilon}{(\hbar \upsilon \vec{G})^{2}}\sum_{j=1}^{3}\left[\operatorname{Re}M_{j,0}\left(\operatorname{Im}\vec{M}_{j}\times\vec{G}_{j}\right)\cdot\hat{z}-\right.\nonumber\\
& \qquad \left.-\operatorname{Im}M_{j}^{0}\left(\operatorname{Re}\vec{M}_{j}\times\vec{G}_{j}\right)\cdot\hat{z}\right]\label{eqn:heffz}
\end{align}
The sums in the above equations are restricted to $j=1,2,3$ due to the relation $\vec{G}_{j+3}=-\vec{G}_{j}$ 
and the property of the corresponding matrices, $M_{j+3}=M_{j}^{\dagger}$. 
We will now prove that $h_{x}=h_{y}=0$. 
The moir\'e Hamiltonian has the property $M_{1}^{0}=M_{3}^{0}=M_{2}^{0*}$ and $M_{1}^{z}=M_{3}^{z}=M_{2}^{z*}$. 
Therefore $\operatorname{Im}M_{3}^{0}M_{3}^{z}=-\operatorname{Im}M_{1}^{0}M_{1}^{z}$. 
However, $\vec{G}_{1}-\vec{G}_{2}+\vec{G}_{3}=0$. 
Examining the above equations, we see that $H_{\rm eff}^{x}=H_{\rm eff}^{y}=0$ due to the symmetry properties of the Hamiltonian, 
and the gap for the unrelaxed configuration arises entirely from a mass term $H_{\rm eff}^{z}$. 

We have numerically calculated the low energy eigenvalues as a function of the parameters $M_{j}^{\alpha}$ to verify the second order perturbation theory result and show which terms contribute at third order and higher. 
Our numerical calculations are performed by multiplying each of the $M_{i}^{\alpha}$ by interpolation parameters $\lambda_{\alpha}$ which range from 0 to 1, thus keeping the same relationship (magnitude and phase) between the different $\vec{G}_{i}$ terms while allowing us to see explicitly the power law behavior of the gap due to each term.

\begin{figure}
\includegraphics[width=8.4cm,angle=0]{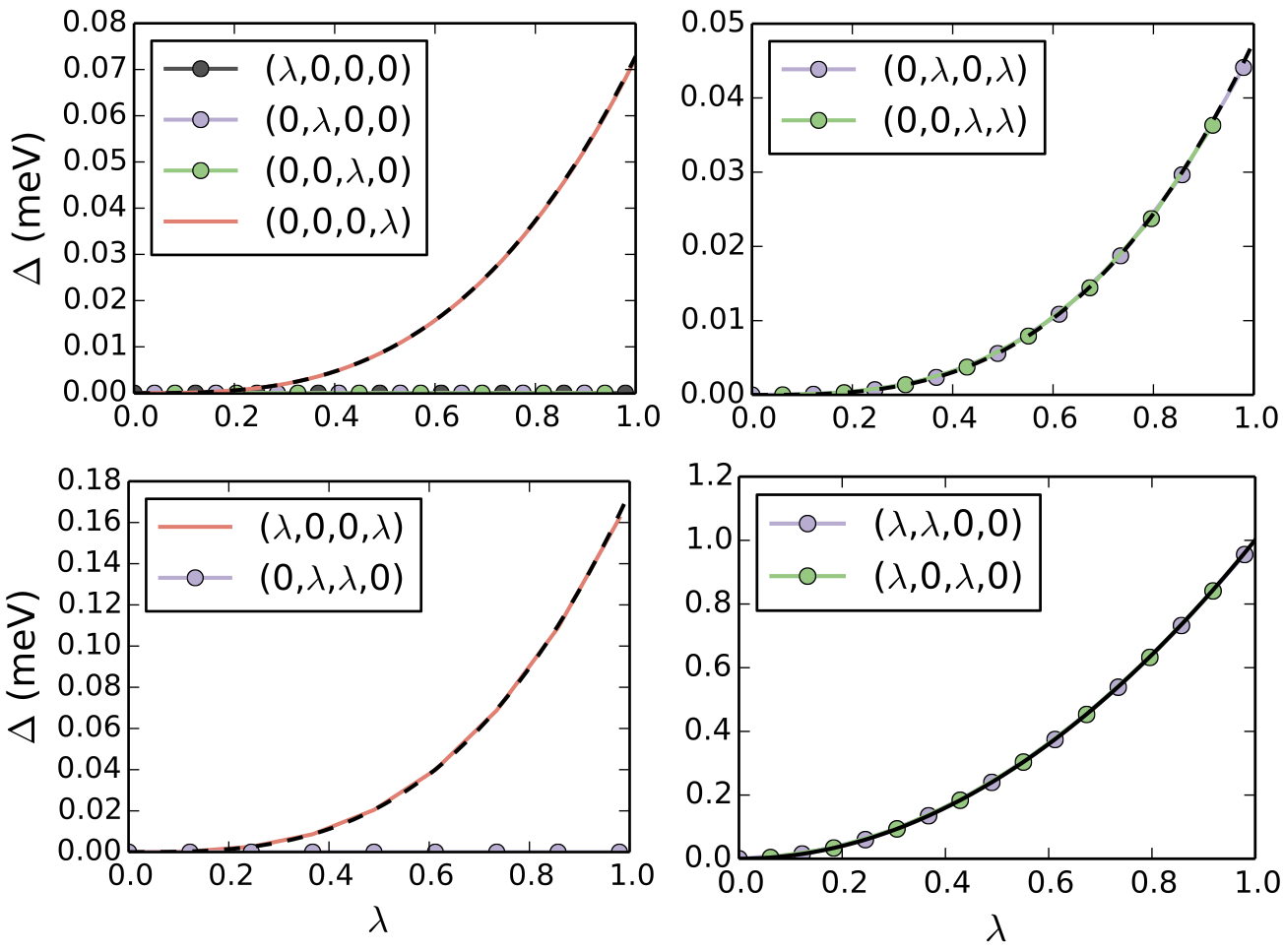}
\caption{{\em Top left:} Gap vs $\lambda$ for individual contributions to $H_{M, \vec{G}}$ from $M^{0}$ (black dots), $M^{x}$ (blue dots), $M^{y}$ (green dots) and $M^{z}$ (red solid line). The labels in the legend correspond to $(\lambda_{0},\,\lambda_{x},\,\lambda_{y},\,\lambda_{z})$. The dashed line is a fit to $\lambda^{3}$.
{\em Top right:} 
Gap vs $\lambda$ for contributions with both $M^{z}$ and either $M^{x}$ (blue) or $M^{y}$ (green). The two terms are equal. Such terms contribute to $H_{\rm eff}^{0}$ at second order (see main text), but $H_{\rm eff}^{0}$ does not contribute to a gap opening. Dashed line (black) is a fit to $\lambda^{3}$, showing that indeed, no second order contribution is evident.
{\em Bottom left:}
Gap vs $\lambda$ for contributions with both $M^{0}$ and $M^{z}$ (solid red line). Such terms contribute to $H_{\rm eff}^{x}$ and $H_{\rm eff}^{y}$ at second order (see main text), which is zero due to the symmetry of the Hamiltonian. Dashed line (black) is a fit to $\lambda^{3}$, showing that indeed, no second order contribution is evident. Also shown (blue circles) is the contribution with both $M^{x}$ and $M^{y}$, which is zero.
{\em Bottom right:}
Gap vs $\lambda$ for contributions with both $M^{0}$ and either $M^{x}$ (blue circles) or $M^{y}$ (green circles). The two contributions are equal. Such terms contribute to $h^{z}$ in the second order perturbation theory (see main text), and therefore contribute to the gap. Solid black line is a fit to $\lambda^{2}$, confirming the perturbation theory result.
}\label{SI:individual}
\end{figure}

First we set all $\lambda_{\alpha}=0$ except for one. Power law fits show that there is no 2nd order contribution from any of the terms individually (Fig.~\ref{SI:individual}). The $\lambda_{z}\neq 0$ term contributes at the 3rd order, while all others are 5th order or higher.

Next we look at the interplay between the different matrix elements $M_{j}^{\alpha}$ which are found to contribute 
to the perturbation theory results for $H_{\rm eff}=H_{\rm eff}^{0}+\vec{H}_{\rm eff}\cdot\vec{\tau}$ as described in the main text. 
Figure \ref{SI:individual} confirm that to second order, the gap is not opened by terms proportional to $M^{x}M^{z}$, $M^{y}M^{z}$, $M^{0}M^{z}$, and $M^{x}M^{y}$. 
The first two, $M^{x}M^{z}$ and $M^{y}M^{z}$ do contribute to the energy levels at second order: 
they lead to a nonzero $H_{\rm eff}^{0}$ which does not open a gap. The $M^{0}M^{z}$ term we found to be zero 
in the second order perturbation theory due to the symmetry of the Hamiltonian, which is verified here. 
Finally, the term $M^{x}M^{y}$ does not appear in the second order perturbation theory at all, 
which is again confirmed by our numerical results. The only terms which contribute to the gap at second order, 
and are therefore most efficient at opening a gap, are $M^{0}M^{x}$ and $M^{0}M^{y}$.

\subsection{Relaxed configuration}
We showed in the main text that the in-plane relaxation of the graphene and boron nitride lattices has a large effect on the size of the gap. 
This is due primarily to the emergence of nonzero mass in the $2\times 2$ block $H^{z}_{M, \vec{G}_0=0}$. This term alone slightly overestimates the gap. We again calculate a second order perturbation theory, Eqn.~\eqref{eqn:Heffective}. However, it is no longer a good approximation to restrict to the nearest six reciprocal lattice vectors. This means that although the decompositions given in Equations~(\ref{eqn:heff0}$-$\ref{eqn:heffz}) remain valid, the symmetry properties that cause $H_{\rm eff}^{x}$ and $H_{\rm eff}^{y}$ to vanish do not strictly hold. We do, however, find that these terms are small. 
The primary contribution to the gap comes from the $H^{z}_{M, \vec{G}_0=0}$ term which overshoots the gap. 
Including the second order terms produces an excellent approximation to the calculated gap. 
Thus we see explicitly that the relaxation is a key source of the gap opening in graphene/boron nitride bilayer systems.


\begin{thebibliography}{10}
\expandafter\ifx\csname url\endcsname\relax
  \def\url#1{\texttt{#1}}\fi
\expandafter\ifx\csname urlprefix\endcsname\relax\def\urlprefix{URL }\fi
\providecommand{\bibinfo}[2]{#2}
\providecommand{\eprint}[2][]{\url{#2}}


\bibitem{dean_seminal}
\bibinfo{author}{Dean, C.~R.} \emph{et~al.}
\newblock \bibinfo{title}{Boron nitride substrates for high-quality graphene
  electronics}.
\newblock \emph{\bibinfo{journal}{Nature Nanotechnology}}
  \textbf{\bibinfo{volume}{5}}, \bibinfo{pages}{722--726}
  (\bibinfo{year}{2010}).


\bibitem{growthgbn}
\bibinfo{author}{Yang, W.} \emph{et~al.}
\newblock \bibinfo{title}{
Epitaxial growth of single-domain graphene on hexagonal boron nitride
}.
\newblock \emph{\bibinfo{journal}{Nature Materials}}
  \textbf{\bibinfo{volume}{12}}, \bibinfo{pages}{792--797}
  (\bibinfo{year}{2013}).



\bibitem{jarillo}
\bibinfo{author}{Hunt, B.} \emph{et~al.}
\newblock \bibinfo{title}{Massive dirac fermions and hofstadter butterfly in a
  van der waals heterostructure}.
\newblock \emph{\bibinfo{journal}{Science}} \textbf{\bibinfo{volume}{340}},
  \bibinfo{pages}{1427--1430} (\bibinfo{year}{2013}).


\bibitem{geimhofstadter}
\bibinfo{author}{Ponomarenko, L.~A.} \emph{et~al.}
\newblock \bibinfo{title}{Cloning of dirac fermions in graphene superlattices}.
\newblock \emph{\bibinfo{journal}{Nature}} \textbf{\bibinfo{volume}{497}},
  \bibinfo{pages}{594--597} (\bibinfo{year}{2013}).


\bibitem{kimhofstadter}
\bibinfo{author}{Dean, C.~R.} \emph{et~al.}
\newblock \bibinfo{title}{Hofstadter's butterfly and the fractal quantum hall
  effect in moir\'e  superlattices}.
\newblock \emph{\bibinfo{journal}{Nature}} \textbf{\bibinfo{volume}{497}},
  \bibinfo{pages}{598--602} (\bibinfo{year}{2013}).


  \bibitem{geimvertical}
\bibinfo{author}{Britnell, L.} \emph{et~al.}
\newblock \bibinfo{title}{Field-Effect Tunneling Transistor Based on Vertical Graphene Heterostructures}.
\newblock \emph{\bibinfo{journal}{Science}} \textbf{\bibinfo{volume}{335}}, 
  \bibinfo{pages}{947--950} (\bibinfo{year}{2012}).
  

\bibitem{leroy}
\bibinfo{author}{Yankowitz, M.} \emph{et~al.}
\newblock \bibinfo{title}{Emergence of superlattice dirac points in graphene on
  hexagonal boron nitride}.
\newblock \emph{\bibinfo{journal}{Nature Physics}}
  \textbf{\bibinfo{volume}{8}}, \bibinfo{pages}{382--386}
  (\bibinfo{year}{2012}).
  
  
  
\bibitem{kelly1}
\bibinfo{author}{Giovannetti, G.}, \bibinfo{author}{Khomyakov, P.~A.},
  \bibinfo{author}{Brocks, G.}, \bibinfo{author}{Kelly, P.~J.} \&
  \bibinfo{author}{van~den Brink, J.}
\newblock \bibinfo{title}{Substrate-induced band gap in graphene on hexagonal
  boron nitride: Ab initio density functional calculations}.
\newblock \emph{\bibinfo{journal}{Physical Review B}}
  \textbf{\bibinfo{volume}{76}}, \bibinfo{pages}{073103}
  (\bibinfo{year}{2007}).
   
  
  \bibitem{moirebandtheory}
\bibinfo{author}{Jung, J.}, \bibinfo{author}{Raoux, A.}, \bibinfo{author}{Qiao,
  Z.} \& \bibinfo{author}{{MacDonald}, A.~H.}
\newblock \bibinfo{title}{Ab-initio theory of moir\'e  superlattice bands in
  layered two-dimensional materials}.
\newblock \emph{\bibinfo{journal}{{Physical  Review B}}}
  \textbf{\bibinfo{volume}{89}}, \bibinfo{pages}{205414}
  (\bibinfo{year}{2014}).
   
   
       \bibitem{ortix}
\bibinfo{author}{Ortix, C.}, \bibinfo{author}{Yang, L.}, \bibinfo{author}{van den Brink, J.}
\newblock \bibinfo{title}{Graphene on incommensurate substrates: Trigonal warping and emerging Dirac cone
replicas with halved group velocity}.
\newblock \emph{\bibinfo{journal}{{Physical  Review B} [cond-mat]}}
  \textbf{\bibinfo{volume}{86}}, \bibinfo{pages}{081405}
  (\bibinfo{year}{2012}).
  

   
\bibitem{geimgap}
\bibinfo{author}{Woods, C.~R.} \emph{et~al.}
\newblock \bibinfo{title}{
Commensurate-incommensurate transition for graphene
  on hexagonal boron nitride}.
\newblock \emph{\bibinfo{journal}{{Nature Physics, DOI:10.1038/nphys2954}}}
  (\bibinfo{year}{2014}).
  
  
     \bibitem{mindthegap}
\bibinfo{author}{Novoselov, K.}
\newblock \bibinfo{title}{Graphene: Mind the gap}.
\newblock \emph{\bibinfo{journal}{Nature Materials}}
  \textbf{\bibinfo{volume}{6}}, \bibinfo{pages}{720--721}
  (\bibinfo{year}{2007}).
  










  
  






\bibitem{falko1}
\bibinfo{author}{Wallbank, J.~R.}, \bibinfo{author}{Patel, A.~A.},
  \bibinfo{author}{Mucha-Kruczy{\'n}ski, M.}, \bibinfo{author}{Geim, A.~K.} \&
  \bibinfo{author}{Fal'ko, V.~I.}
\newblock \bibinfo{title}{Generic miniband structure of graphene on a hexagonal
  substrate}.
\newblock \emph{\bibinfo{journal}{Physical Review B}}
  \textbf{\bibinfo{volume}{87}}, \bibinfo{pages}{245408}
  (\bibinfo{year}{2013}).


\bibitem{alignment}
\bibinfo{author}{Tang, S.} \emph{et~al.}
\newblock \bibinfo{title}{Precisely aligned graphene grown on hexagonal boron
  nitride by catalyst free chemical vapor deposition}.
\newblock \emph{\bibinfo{journal}{Scientific Reports}}
  \textbf{\bibinfo{volume}{3}} (\bibinfo{year}{2013}).



\bibitem{pedersen}
\bibinfo{author}{Pedersen, T.~G.} \emph{et~al.}
\newblock \bibinfo{title}{Graphene antidot lattices: Designed defects and spin
  qubits}.
\newblock \emph{\bibinfo{journal}{Physical Review Letters}}
  \textbf{\bibinfo{volume}{100}}, \bibinfo{pages}{136804}
  (\bibinfo{year}{2008}).

\bibitem{snyman}
\bibinfo{author}{Snyman, I.}
\newblock \bibinfo{title}{Gapped state of a carbon monolayer in periodic
  magnetic and electric fields}.
\newblock \emph{\bibinfo{journal}{Physical Review B}}
  \textbf{\bibinfo{volume}{80}}, \bibinfo{pages}{054303}
  (\bibinfo{year}{2009}).

\bibitem{guinealow}
\bibinfo{author}{Low, T.}, \bibinfo{author}{Guinea, F.} \&
  \bibinfo{author}{Katsnelson, M.~I.}
\newblock \bibinfo{title}{Gaps tunable by electrostatic gates in strained
  graphene}.
\newblock \emph{\bibinfo{journal}{Physical Review B}}
  \textbf{\bibinfo{volume}{83}}, \bibinfo{pages}{195436}
  (\bibinfo{year}{2011}).

\bibitem{kindermann}
\bibinfo{author}{Kindermann, M.}, \bibinfo{author}{Uchoa, B.} \&
  \bibinfo{author}{Miller, D.~L.}
\newblock \bibinfo{title}{Zero-energy modes and gate-tunable gap in graphene on
  hexagonal boron nitride}.
\newblock \emph{\bibinfo{journal}{Physical Review B}}
  \textbf{\bibinfo{volume}{86}}, \bibinfo{pages}{115415}
  (\bibinfo{year}{2012}).
  
  \bibitem{rafiallan}
\bibinfo{author}{Bistritzer, R.} \& \bibinfo{author}{{MacDonald}, A.~H.}
\newblock \bibinfo{title}{Moir\'e bands in twisted double-layer graphene}.
\newblock \emph{\bibinfo{journal}{Proceedings of the National Academy of
  Sciences}} \textbf{\bibinfo{volume}{108}}, \bibinfo{pages}{12233--12237}
  (\bibinfo{year}{2011}).
  

\bibitem{vozmedianopseudo}
\bibinfo{author}{Vozmediano, M.}, \bibinfo{author}{Katsnelson, M.} \&
  \bibinfo{author}{Guinea, F.}
\newblock \bibinfo{title}{Gauge fields in graphene}.
\newblock \emph{\bibinfo{journal}{Physics Reports}}
  \textbf{\bibinfo{volume}{496}}, \bibinfo{pages}{109--148}
  (\bibinfo{year}{2010}).
  
  \bibitem{physicascripta} 
  \bibinfo{author}{MacDonald, A. H. }, 
  \bibinfo{author}{Jung, J.},
  \bibinfo{author}{Zhang, F.},
  \newblock \bibinfo{title}{Pseudospin Order in Monolayer, Bilayer, and Double-Layer Graphene}.
\newblock \emph{\bibinfo{journal}{Phys. Scr.}}
\bibinfo{pages}{014012}
  (\bibinfo{year}{2012}).

 
  
  
  \bibitem{kellygap}
\bibinfo{author}{Bokdam M.} \bibinfo{author}{Amlaki T.} \bibinfo{author}{Kelly P. J.}
\newblock \bibinfo{title}{Band gaps in incommensurable graphene on hexagonal boron nitride},
\newblock \emph{\bibinfo{journal}{Physical Review B}}
\textbf{\bibinfo{volume}{89}} \bibinfo{pages}{201404(R)} (\bibinfo{year}{2014}).



\bibitem{song}
Song, J. C. W.,  Shytov, A. V., and Levitov, L. S., 
Electron interactions and gap opening in graphene superlattices,
{\em Physical Review Letters} {\bf 111}, 266801 (2013).



\bibitem{trilayergap}
Jung J., and MacDonald, A. H. 
Gapped broken symmetry states in ABC-stacked trilayer graphene,
Physical Review B {\bf 88}, 075408 (2013).
   

\bibitem{moireHF}
DaSilva A. M. {\em et al.} unpublished.



\bibitem{sachs}
\bibinfo{author}{Sachs, B.}, \bibinfo{author}{Wehling, T.~O.},
  \bibinfo{author}{Katsnelson, M.~I.} \& \bibinfo{author}{Lichtenstein, A.~I.}
\newblock \bibinfo{title}{Adhesion and electronic structure of graphene on
  hexagonal boron nitride substrates}.
\newblock \emph{\bibinfo{journal}{Physical Review B}}
  \textbf{\bibinfo{volume}{84}}, \bibinfo{pages}{195414}
  (\bibinfo{year}{2011}).


 
   
   
   
\bibitem{elasticitybook}
\bibinfo{author}{Washizu, K.}
\newblock \emph{\bibinfo{title}{Variational methods in elasticity and
  plasticity}} (\bibinfo{publisher}{Pergamon Press}, \bibinfo{year}{1975}).

\bibitem{bendingstiffness}
\bibinfo{author}{Koskinen, P.} \& \bibinfo{author}{Kit, O.~O.}
\newblock \bibinfo{title}{Approximate modeling of spherical membranes}.
\newblock \emph{\bibinfo{journal}{Physical Review B}}
  \textbf{\bibinfo{volume}{82}}, \bibinfo{pages}{235420}
  (\bibinfo{year}{2010}).

\bibitem{katsnelson}
\bibinfo{author}{Zakharchenko, K.~V.}, \bibinfo{author}{Katsnelson, M.~I.} \&
  \bibinfo{author}{Fasolino, A.}
\newblock \bibinfo{title}{Finite temperature lattice properties of graphene
  beyond the quasiharmonic approximation}.
\newblock \emph{\bibinfo{journal}{Physical Review Letters}}
  \textbf{\bibinfo{volume}{102}}, \bibinfo{pages}{046808}
  (\bibinfo{year}{2009}).



   
   \bibitem{graphenetb}
\bibinfo{author}{Jung, J.}, \bibinfo{author}{MacDonald, A.~H.}
\newblock \bibinfo{title}{Tight-binding model for graphene $\pi$-bands from maximally localized Wannier functions},
\newblock \emph{\bibinfo{journal}{Physical Review B}}
  \textbf{\bibinfo{volume}{87}} \bibinfo{pages}{195450} (\bibinfo{year}{2013}).
\end{thebibliography}

\begin{thebibliography}{10}
\expandafter\ifx\csname url\endcsname\relax
  \def\url#1{\texttt{#1}}\fi
\expandafter\ifx\csname urlprefix\endcsname\relax\def\urlprefix{URL }\fi
\providecommand{\bibinfo}[2]{#2}
\providecommand{\eprint}[2][]{\url{#2}}

\bibitem{supp_moirebandtheory}
\bibinfo{author}{Jung, J.}, \bibinfo{author}{Raoux, A.}, 
\bibinfo{author}{Qiao,  Z.} \& \bibinfo{author}{{MacDonald}, A.~H.}
\newblock \bibinfo{title}{Ab-initio theory of moire superlattice bands in
  layered two-dimensional materials}.
\newblock \emph{\bibinfo{journal}{Physical  Review B}}
  \textbf{\bibinfo{volume}{89}}, \bibinfo{pages}{205414}
  (\bibinfo{year}{2014}).


\bibitem{supp_graphenetb}
\bibinfo{author}{Jung, J.} \& \bibinfo{author}{{MacDonald}, A.~H.}
\newblock \bibinfo{title}{Tight-binding model for graphene $\pi$€-bands from
  maximally localized wannier functions}.
\newblock \emph{\bibinfo{journal}{Physical Review B}}
  \textbf{\bibinfo{volume}{87}}, \bibinfo{pages}{195450}
  (\bibinfo{year}{2013}).


\bibitem{supp_bendingstiffness}
\bibinfo{author}{Koskinen, P.} \& \bibinfo{author}{Kit, O.~O.}
\newblock \bibinfo{title}{Approximate modeling of spherical membranes}.
\newblock \emph{\bibinfo{journal}{Physical Review B}}
  \textbf{\bibinfo{volume}{82}}, \bibinfo{pages}{235420}
  (\bibinfo{year}{2010}).


\bibitem{supp_katsnelson}
\bibinfo{author}{Zakharchenko, K.~V.}, \bibinfo{author}{Katsnelson, M.~I.} \&
  \bibinfo{author}{Fasolino, A.}
\newblock \bibinfo{title}{Finite temperature lattice properties of graphene
  beyond the quasiharmonic approximation}.
\newblock \emph{\bibinfo{journal}{Physical Review Letters}}
  \textbf{\bibinfo{volume}{102}}, \bibinfo{pages}{046808}
  (\bibinfo{year}{2009}).
\newblock


\bibitem{supp_elasticitybook}
\bibinfo{author}{Washizu, K.}
\newblock \emph{\bibinfo{title}{Variational methods in elasticity and
  plasticity}} (\bibinfo{publisher}{Pergamon Press}, \bibinfo{year}{1975}).

\bibitem{supp_guinearipples}
\bibinfo{author}{Guinea, F.}, \bibinfo{author}{Horovitz, B.} \&
  \bibinfo{author}{Le~Doussal, P.}
\newblock \bibinfo{title}{Gauge field induced by ripples in graphene}.
\newblock \emph{\bibinfo{journal}{Physical Review B}}
  \textbf{\bibinfo{volume}{77}}, \bibinfo{pages}{205421}
  (\bibinfo{year}{2008}).


\bibitem{supp_sachs}
\bibinfo{author}{Sachs, B.}, \bibinfo{author}{Wehling, T.~O.},
  \bibinfo{author}{Katsnelson, M.~I.} \& \bibinfo{author}{Lichtenstein, A.~I.}
\newblock \bibinfo{title}{Adhesion and electronic structure of graphene on
  hexagonal boron nitride substrates}.
\newblock \emph{\bibinfo{journal}{Physical Review B}}
  \textbf{\bibinfo{volume}{84}}, \bibinfo{pages}{195414}
  (\bibinfo{year}{2011}).


\bibitem{supp_frenkelkontorovabook}
\bibinfo{author}{Braun, O.~M.} \& \bibinfo{author}{Kivshar, Y.~S.}
\newblock \emph{\bibinfo{title}{The Frenkel-Kontorova Model}}
  (\bibinfo{publisher}{Springer-Verlag}, \bibinfo{year}{2004}).

\bibitem{supp_gould}
\bibinfo{author}{Gould, T.}, \bibinfo{author}{Leb{\`e}gue, S.} \&
  \bibinfo{author}{Dobson, J.~F.}
\newblock \bibinfo{title}{Dispersion corrections in graphenic systems: a simple
  and effective model of binding}.
\newblock \emph{\bibinfo{journal}{Journal of Physics: Condensed Matter}}
  \textbf{\bibinfo{volume}{25}}, \bibinfo{pages}{445010}
  (\bibinfo{year}{2013}).
\newblock


\bibitem{supp_apstalk}
\bibinfo{title}{Bull. Am. Phys. Soc. 59, 1 455 (2014). http://meetings.aps.org/link/BAPS.2014.MAR.F55.3}

\bibitem{supp_jelliumrpa}
\bibinfo{author}{Jung, J.}, \bibinfo{author}{Garc{\'\i}a-Gonz{\'a}lez, P.},
  \bibinfo{author}{Dobson, J.~F.} \& \bibinfo{author}{Godby, R.~W.}
\newblock \bibinfo{title}{Effects beyond the random-phase approximation in
  calculating the interaction between metal films}.
\newblock \emph{\bibinfo{journal}{Physical Review B}}
  \textbf{\bibinfo{volume}{70}}, \bibinfo{pages}{205107}
  (\bibinfo{year}{2004}).


\bibitem{supp_hbnacfd}
\bibinfo{author}{Marini, A.}, \bibinfo{author}{Garc{\'\i}a-Gonz{\'a}lez, P.} \&
  \bibinfo{author}{Rubio, A.}
\newblock \bibinfo{title}{First-principles description of correlation effects
  in layered materials}.
\newblock \emph{\bibinfo{journal}{Physical Review Letters}}
  \textbf{\bibinfo{volume}{96}}, \bibinfo{pages}{136404}
  (\bibinfo{year}{2006}).
\newblock


\bibitem{supp_graphitebinding}
\bibinfo{author}{Leb{\`e}gue, S.} \emph{et~al.}
\newblock \bibinfo{title}{Cohesive properties and asymptotics of the dispersion
  interaction in graphite by the random phase approximation}.
\newblock \emph{\bibinfo{journal}{Physical Review Letters}}
  \textbf{\bibinfo{volume}{105}}, \bibinfo{pages}{196401}
  (\bibinfo{year}{2010}).
\newblock


\bibitem{supp_bjorkman}
\bibinfo{author}{Bj{\"o}rkman, T.}, \bibinfo{author}{Gulans, A.},
  \bibinfo{author}{Krasheninnikov, A.~V.} \& \bibinfo{author}{Nieminen, R.~M.}
\newblock \bibinfo{title}{van der waals bonding in layered compounds from
  advanced density-functional first-principles calculations}.
\newblock \emph{\bibinfo{journal}{Physical Review Letters}}
  \textbf{\bibinfo{volume}{108}}, \bibinfo{pages}{235502}
  (\bibinfo{year}{2012}).
\newblock


\bibitem{supp_marom}
\bibinfo{author}{Marom, N.} \emph{et~al.}
\newblock \bibinfo{title}{Stacking and registry effects in layered materials:
  The case of hexagonal boron nitride}.
\newblock \emph{\bibinfo{journal}{Physical Review Letters}}
  \textbf{\bibinfo{volume}{105}}, \bibinfo{pages}{046801}
  (\bibinfo{year}{2010}).
\newblock


 \bibitem{supp_suzuura}
 \bibinfo{author}{Suzuura H.} \& \bibinfo{author}{Ando T.} \&
\newblock \bibinfo{title}{Phonons and electron-phonon scattering in carbon nanotubes}.
\newblock \emph{\bibinfo{journal}{Physical Review B}}
  \textbf{\bibinfo{volume}{65}}, \bibinfo{pages}{235412}
  (\bibinfo{year}{2002}).
\newblock
%
 
 
\bibitem{supp_vozmedianopseudo}
\bibinfo{author}{Vozmediano, M.}, \bibinfo{author}{Katsnelson, M.} \&
  \bibinfo{author}{Guinea, F.}
\newblock \bibinfo{title}{Gauge fields in graphene}.
\newblock \emph{\bibinfo{journal}{Physics Reports}}
  \textbf{\bibinfo{volume}{496}}, \bibinfo{pages}{109--148}
  (\bibinfo{year}{2010}).
\newblock

  
     
 
\end{thebibliography}
\end{document}